\documentstyle[graphicx,psfig,epsfig,amsmath,amssymb,enumerate,color]{mn}

\definecolor{AliceBlue}{rgb}{0.94,0.97,1.00}
\definecolor{AntiqueWhite1}{rgb}{1.00,0.94,0.86}
\definecolor{AntiqueWhite2}{rgb}{0.93,0.87,0.80}
\definecolor{AntiqueWhite3}{rgb}{0.80,0.75,0.69}
\definecolor{AntiqueWhite4}{rgb}{0.55,0.51,0.47}
\definecolor{AntiqueWhite}{rgb}{0.98,0.92,0.84}
\definecolor{BlanchedAlmond}{rgb}{1.00,0.92,0.80}
\definecolor{BlueViolet}{rgb}{0.54,0.17,0.89}
\definecolor{CadetBlue1}{rgb}{0.60,0.96,1.00}
\definecolor{CadetBlue2}{rgb}{0.56,0.90,0.93}
\definecolor{CadetBlue3}{rgb}{0.48,0.77,0.80}
\definecolor{CadetBlue4}{rgb}{0.33,0.53,0.55}
\definecolor{CadetBlue}{rgb}{0.37,0.62,0.63}
\definecolor{CornflowerBlue}{rgb}{0.39,0.58,0.93}
\definecolor{DarkBlue}{rgb}{0.00,0.00,0.55}
\definecolor{DarkCyan}{rgb}{0.00,0.55,0.55}
\definecolor{DarkGoldenrod1}{rgb}{1.00,0.73,0.06}
\definecolor{DarkGoldenrod2}{rgb}{0.93,0.68,0.05}
\definecolor{DarkGoldenrod3}{rgb}{0.80,0.58,0.05}
\definecolor{DarkGoldenrod4}{rgb}{0.55,0.40,0.03}
\definecolor{DarkGoldenrod}{rgb}{0.72,0.53,0.04}
\definecolor{DarkGray}{rgb}{0.66,0.66,0.66}
\definecolor{DarkGreen}{rgb}{0.00,0.39,0.00}
\definecolor{DarkGrey}{rgb}{0.66,0.66,0.66}
\definecolor{DarkKhaki}{rgb}{0.74,0.72,0.42}
\definecolor{DarkMagenta}{rgb}{0.55,0.00,0.55}
\definecolor{DarkOliveGreen1}{rgb}{0.79,1.00,0.44}
\definecolor{DarkOliveGreen2}{rgb}{0.74,0.93,0.41}
\definecolor{DarkOliveGreen3}{rgb}{0.64,0.80,0.35}
\definecolor{DarkOliveGreen4}{rgb}{0.43,0.55,0.24}
\definecolor{DarkOliveGreen}{rgb}{0.33,0.42,0.18}
\definecolor{DarkOrange1}{rgb}{1.00,0.50,0.00}
\definecolor{DarkOrange2}{rgb}{0.93,0.46,0.00}
\definecolor{DarkOrange3}{rgb}{0.80,0.40,0.00}
\definecolor{DarkOrange4}{rgb}{0.55,0.27,0.00}
\definecolor{DarkOrange}{rgb}{1.00,0.55,0.00}
\definecolor{DarkOrchid1}{rgb}{0.75,0.24,1.00}
\definecolor{DarkOrchid2}{rgb}{0.70,0.23,0.93}
\definecolor{DarkOrchid3}{rgb}{0.60,0.20,0.80}
\definecolor{DarkOrchid4}{rgb}{0.41,0.13,0.55}
\definecolor{DarkOrchid}{rgb}{0.60,0.20,0.80}
\definecolor{DarkRed}{rgb}{0.55,0.00,0.00}
\definecolor{DarkSalmon}{rgb}{0.91,0.59,0.48}
\definecolor{DarkSeaGreen1}{rgb}{0.76,1.00,0.76}
\definecolor{DarkSeaGreen2}{rgb}{0.71,0.93,0.71}
\definecolor{DarkSeaGreen3}{rgb}{0.61,0.80,0.61}
\definecolor{DarkSeaGreen4}{rgb}{0.41,0.55,0.41}
\definecolor{DarkSeaGreen}{rgb}{0.56,0.74,0.56}
\definecolor{DarkSlateBlue}{rgb}{0.28,0.24,0.55}
\definecolor{DarkSlateGray1}{rgb}{0.59,1.00,1.00}
\definecolor{DarkSlateGray2}{rgb}{0.55,0.93,0.93}
\definecolor{DarkSlateGray3}{rgb}{0.47,0.80,0.80}
\definecolor{DarkSlateGray4}{rgb}{0.32,0.55,0.55}
\definecolor{DarkSlateGray}{rgb}{0.18,0.31,0.31}
\definecolor{DarkSlateGrey}{rgb}{0.18,0.31,0.31}
\definecolor{DarkTurquoise}{rgb}{0.00,0.81,0.82}
\definecolor{DarkViolet}{rgb}{0.58,0.00,0.83}
\definecolor{DeepPink1}{rgb}{1.00,0.08,0.58}
\definecolor{DeepPink2}{rgb}{0.93,0.07,0.54}
\definecolor{DeepPink3}{rgb}{0.80,0.06,0.46}
\definecolor{DeepPink4}{rgb}{0.55,0.04,0.31}
\definecolor{DeepPink}{rgb}{1.00,0.08,0.58}
\definecolor{DeepSkyBlue1}{rgb}{0.00,0.75,1.00}
\definecolor{DeepSkyBlue2}{rgb}{0.00,0.70,0.93}
\definecolor{DeepSkyBlue3}{rgb}{0.00,0.60,0.80}
\definecolor{DeepSkyBlue4}{rgb}{0.00,0.41,0.55}
\definecolor{DeepSkyBlue}{rgb}{0.00,0.75,1.00}
\definecolor{DimGray}{rgb}{0.41,0.41,0.41}
\definecolor{DimGrey}{rgb}{0.41,0.41,0.41}
\definecolor{DodgerBlue1}{rgb}{0.12,0.56,1.00}
\definecolor{DodgerBlue2}{rgb}{0.11,0.53,0.93}
\definecolor{DodgerBlue3}{rgb}{0.09,0.45,0.80}
\definecolor{DodgerBlue4}{rgb}{0.06,0.31,0.55}
\definecolor{DodgerBlue}{rgb}{0.12,0.56,1.00}
\definecolor{FloralWhite}{rgb}{1.00,0.98,0.94}
\definecolor{ForestGreen}{rgb}{0.13,0.55,0.13}
\definecolor{GhostWhite}{rgb}{0.97,0.97,1.00}
\definecolor{GreenYellow}{rgb}{0.68,1.00,0.18}
\definecolor{HotPink1}{rgb}{1.00,0.43,0.71}
\definecolor{HotPink2}{rgb}{0.93,0.42,0.65}
\definecolor{HotPink3}{rgb}{0.80,0.38,0.56}
\definecolor{HotPink4}{rgb}{0.55,0.23,0.38}
\definecolor{HotPink}{rgb}{1.00,0.41,0.71}
\definecolor{IndianRed1}{rgb}{1.00,0.42,0.42}
\definecolor{IndianRed2}{rgb}{0.93,0.39,0.39}
\definecolor{IndianRed3}{rgb}{0.80,0.33,0.33}
\definecolor{IndianRed4}{rgb}{0.55,0.23,0.23}
\definecolor{IndianRed}{rgb}{0.80,0.36,0.36}
\definecolor{LavenderBlush1}{rgb}{1.00,0.94,0.96}
\definecolor{LavenderBlush2}{rgb}{0.93,0.88,0.90}
\definecolor{LavenderBlush3}{rgb}{0.80,0.76,0.77}
\definecolor{LavenderBlush4}{rgb}{0.55,0.51,0.53}
\definecolor{LavenderBlush}{rgb}{1.00,0.94,0.96}
\definecolor{LawnGreen}{rgb}{0.49,0.99,0.00}
\definecolor{LemonChiffon1}{rgb}{1.00,0.98,0.80}
\definecolor{LemonChiffon2}{rgb}{0.93,0.91,0.75}
\definecolor{LemonChiffon3}{rgb}{0.80,0.79,0.65}
\definecolor{LemonChiffon4}{rgb}{0.55,0.54,0.44}
\definecolor{LemonChiffon}{rgb}{1.00,0.98,0.80}
\definecolor{LightBlue1}{rgb}{0.75,0.94,1.00}
\definecolor{LightBlue2}{rgb}{0.70,0.87,0.93}
\definecolor{LightBlue3}{rgb}{0.60,0.75,0.80}
\definecolor{LightBlue4}{rgb}{0.41,0.51,0.55}
\definecolor{LightBlue}{rgb}{0.68,0.85,0.90}
\definecolor{LightCoral}{rgb}{0.94,0.50,0.50}
\definecolor{LightCyan1}{rgb}{0.88,1.00,1.00}
\definecolor{LightCyan2}{rgb}{0.82,0.93,0.93}
\definecolor{LightCyan3}{rgb}{0.71,0.80,0.80}
\definecolor{LightCyan4}{rgb}{0.48,0.55,0.55}
\definecolor{LightCyan}{rgb}{0.88,1.00,1.00}
\definecolor{LightGoldenrod1}{rgb}{1.00,0.93,0.55}
\definecolor{LightGoldenrod2}{rgb}{0.93,0.86,0.51}
\definecolor{LightGoldenrod3}{rgb}{0.80,0.75,0.44}
\definecolor{LightGoldenrod4}{rgb}{0.55,0.51,0.30}
\definecolor{LightGoldenrodYellow}{rgb}{0.98,0.98,0.82}
\definecolor{LightGoldenrod}{rgb}{0.93,0.87,0.51}
\definecolor{LightGray}{rgb}{0.83,0.83,0.83}
\definecolor{LightGreen}{rgb}{0.56,0.93,0.56}
\definecolor{LightGrey}{rgb}{0.83,0.83,0.83}
\definecolor{LightPink1}{rgb}{1.00,0.68,0.73}
\definecolor{LightPink2}{rgb}{0.93,0.64,0.68}
\definecolor{LightPink3}{rgb}{0.80,0.55,0.58}
\definecolor{LightPink4}{rgb}{0.55,0.37,0.40}
\definecolor{LightPink}{rgb}{1.00,0.71,0.76}
\definecolor{LightSalmon1}{rgb}{1.00,0.63,0.48}
\definecolor{LightSalmon2}{rgb}{0.93,0.58,0.45}
\definecolor{LightSalmon3}{rgb}{0.80,0.51,0.38}
\definecolor{LightSalmon4}{rgb}{0.55,0.34,0.26}
\definecolor{LightSalmon}{rgb}{1.00,0.63,0.48}
\definecolor{LightSeaGreen}{rgb}{0.13,0.70,0.67}
\definecolor{LightSkyBlue1}{rgb}{0.69,0.89,1.00}
\definecolor{LightSkyBlue2}{rgb}{0.64,0.83,0.93}
\definecolor{LightSkyBlue3}{rgb}{0.55,0.71,0.80}
\definecolor{LightSkyBlue4}{rgb}{0.38,0.48,0.55}
\definecolor{LightSkyBlue}{rgb}{0.53,0.81,0.98}
\definecolor{LightSlateBlue}{rgb}{0.52,0.44,1.00}
\definecolor{LightSlateGray}{rgb}{0.47,0.53,0.60}
\definecolor{LightSlateGrey}{rgb}{0.47,0.53,0.60}
\definecolor{LightSteelBlue1}{rgb}{0.79,0.88,1.00}
\definecolor{LightSteelBlue2}{rgb}{0.74,0.82,0.93}
\definecolor{LightSteelBlue3}{rgb}{0.64,0.71,0.80}
\definecolor{LightSteelBlue4}{rgb}{0.43,0.48,0.55}
\definecolor{LightSteelBlue}{rgb}{0.69,0.77,0.87}
\definecolor{LightYellow1}{rgb}{1.00,1.00,0.88}
\definecolor{LightYellow2}{rgb}{0.93,0.93,0.82}
\definecolor{LightYellow3}{rgb}{0.80,0.80,0.71}
\definecolor{LightYellow4}{rgb}{0.55,0.55,0.48}
\definecolor{LightYellow}{rgb}{1.00,1.00,0.88}
\definecolor{LimeGreen}{rgb}{0.20,0.80,0.20}
\definecolor{MediumAquamarine}{rgb}{0.40,0.80,0.67}
\definecolor{MediumBlue}{rgb}{0.00,0.00,0.80}
\definecolor{MediumOrchid1}{rgb}{0.88,0.40,1.00}
\definecolor{MediumOrchid2}{rgb}{0.82,0.37,0.93}
\definecolor{MediumOrchid3}{rgb}{0.71,0.32,0.80}
\definecolor{MediumOrchid4}{rgb}{0.48,0.22,0.55}
\definecolor{MediumOrchid}{rgb}{0.73,0.33,0.83}
\definecolor{MediumPurple1}{rgb}{0.67,0.51,1.00}
\definecolor{MediumPurple2}{rgb}{0.62,0.47,0.93}
\definecolor{MediumPurple3}{rgb}{0.54,0.41,0.80}
\definecolor{MediumPurple4}{rgb}{0.36,0.28,0.55}
\definecolor{MediumPurple}{rgb}{0.58,0.44,0.86}
\definecolor{MediumSeaGreen}{rgb}{0.24,0.70,0.44}
\definecolor{MediumSlateBlue}{rgb}{0.48,0.41,0.93}
\definecolor{MediumSpringGreen}{rgb}{0.00,0.98,0.60}
\definecolor{MediumTurquoise}{rgb}{0.28,0.82,0.80}
\definecolor{MediumVioletRed}{rgb}{0.78,0.08,0.52}
\definecolor{MidnightBlue}{rgb}{0.10,0.10,0.44}
\definecolor{MintCream}{rgb}{0.96,1.00,0.98}
\definecolor{MistyRose1}{rgb}{1.00,0.89,0.88}
\definecolor{MistyRose2}{rgb}{0.93,0.84,0.82}
\definecolor{MistyRose3}{rgb}{0.80,0.72,0.71}
\definecolor{MistyRose4}{rgb}{0.55,0.49,0.48}
\definecolor{MistyRose}{rgb}{1.00,0.89,0.88}
\definecolor{NavajoWhite1}{rgb}{1.00,0.87,0.68}
\definecolor{NavajoWhite2}{rgb}{0.93,0.81,0.63}
\definecolor{NavajoWhite3}{rgb}{0.80,0.70,0.55}
\definecolor{NavajoWhite4}{rgb}{0.55,0.47,0.37}
\definecolor{NavajoWhite}{rgb}{1.00,0.87,0.68}
\definecolor{NavyBlue}{rgb}{0.00,0.00,0.50}
\definecolor{OldLace}{rgb}{0.99,0.96,0.90}
\definecolor{OliveDrab1}{rgb}{0.75,1.00,0.24}
\definecolor{OliveDrab2}{rgb}{0.70,0.93,0.23}
\definecolor{OliveDrab3}{rgb}{0.60,0.80,0.20}
\definecolor{OliveDrab4}{rgb}{0.41,0.55,0.13}
\definecolor{OliveDrab}{rgb}{0.42,0.56,0.14}
\definecolor{OrangeRed1}{rgb}{1.00,0.27,0.00}
\definecolor{OrangeRed2}{rgb}{0.93,0.25,0.00}
\definecolor{OrangeRed3}{rgb}{0.80,0.22,0.00}
\definecolor{OrangeRed4}{rgb}{0.55,0.15,0.00}
\definecolor{OrangeRed}{rgb}{1.00,0.27,0.00}
\definecolor{PaleGoldenrod}{rgb}{0.93,0.91,0.67}
\definecolor{PaleGreen1}{rgb}{0.60,1.00,0.60}
\definecolor{PaleGreen2}{rgb}{0.56,0.93,0.56}
\definecolor{PaleGreen3}{rgb}{0.49,0.80,0.49}
\definecolor{PaleGreen4}{rgb}{0.33,0.55,0.33}
\definecolor{PaleGreen}{rgb}{0.60,0.98,0.60}
\definecolor{PaleTurquoise1}{rgb}{0.73,1.00,1.00}
\definecolor{PaleTurquoise2}{rgb}{0.68,0.93,0.93}
\definecolor{PaleTurquoise3}{rgb}{0.59,0.80,0.80}
\definecolor{PaleTurquoise4}{rgb}{0.40,0.55,0.55}
\definecolor{PaleTurquoise}{rgb}{0.69,0.93,0.93}
\definecolor{PaleVioletRed1}{rgb}{1.00,0.51,0.67}
\definecolor{PaleVioletRed2}{rgb}{0.93,0.47,0.62}
\definecolor{PaleVioletRed3}{rgb}{0.80,0.41,0.54}
\definecolor{PaleVioletRed4}{rgb}{0.55,0.28,0.36}
\definecolor{PaleVioletRed}{rgb}{0.86,0.44,0.58}
\definecolor{PapayaWhip}{rgb}{1.00,0.94,0.84}
\definecolor{PeachPuff1}{rgb}{1.00,0.85,0.73}
\definecolor{PeachPuff2}{rgb}{0.93,0.80,0.68}
\definecolor{PeachPuff3}{rgb}{0.80,0.69,0.58}
\definecolor{PeachPuff4}{rgb}{0.55,0.47,0.40}
\definecolor{PeachPuff}{rgb}{1.00,0.85,0.73}
\definecolor{PowderBlue}{rgb}{0.69,0.88,0.90}
\definecolor{RosyBrown1}{rgb}{1.00,0.76,0.76}
\definecolor{RosyBrown2}{rgb}{0.93,0.71,0.71}
\definecolor{RosyBrown3}{rgb}{0.80,0.61,0.61}
\definecolor{RosyBrown4}{rgb}{0.55,0.41,0.41}
\definecolor{RosyBrown}{rgb}{0.74,0.56,0.56}
\definecolor{RoyalBlue1}{rgb}{0.28,0.46,1.00}
\definecolor{RoyalBlue2}{rgb}{0.26,0.43,0.93}
\definecolor{RoyalBlue3}{rgb}{0.23,0.37,0.80}
\definecolor{RoyalBlue4}{rgb}{0.15,0.25,0.55}
\definecolor{RoyalBlue}{rgb}{0.25,0.41,0.88}
\definecolor{SaddleBrown}{rgb}{0.55,0.27,0.07}
\definecolor{SandyBrown}{rgb}{0.96,0.64,0.38}
\definecolor{SeaGreen1}{rgb}{0.33,1.00,0.62}
\definecolor{SeaGreen2}{rgb}{0.31,0.93,0.58}
\definecolor{SeaGreen3}{rgb}{0.26,0.80,0.50}
\definecolor{SeaGreen4}{rgb}{0.18,0.55,0.34}
\definecolor{SeaGreen}{rgb}{0.18,0.55,0.34}
\definecolor{SkyBlue1}{rgb}{0.53,0.81,1.00}
\definecolor{SkyBlue2}{rgb}{0.49,0.75,0.93}
\definecolor{SkyBlue3}{rgb}{0.42,0.65,0.80}
\definecolor{SkyBlue4}{rgb}{0.29,0.44,0.55}
\definecolor{SkyBlue}{rgb}{0.53,0.81,0.92}
\definecolor{SlateBlue1}{rgb}{0.51,0.44,1.00}
\definecolor{SlateBlue2}{rgb}{0.48,0.40,0.93}
\definecolor{SlateBlue3}{rgb}{0.41,0.35,0.80}
\definecolor{SlateBlue4}{rgb}{0.28,0.24,0.55}
\definecolor{SlateBlue}{rgb}{0.42,0.35,0.80}
\definecolor{SlateGray1}{rgb}{0.78,0.89,1.00}
\definecolor{SlateGray2}{rgb}{0.73,0.83,0.93}
\definecolor{SlateGray3}{rgb}{0.62,0.71,0.80}
\definecolor{SlateGray4}{rgb}{0.42,0.48,0.55}
\definecolor{SlateGray}{rgb}{0.44,0.50,0.56}
\definecolor{SlateGrey}{rgb}{0.44,0.50,0.56}
\definecolor{SpringGreen1}{rgb}{0.00,1.00,0.50}
\definecolor{SpringGreen2}{rgb}{0.00,0.93,0.46}
\definecolor{SpringGreen3}{rgb}{0.00,0.80,0.40}
\definecolor{SpringGreen4}{rgb}{0.00,0.55,0.27}
\definecolor{SpringGreen}{rgb}{0.00,1.00,0.50}
\definecolor{SteelBlue1}{rgb}{0.39,0.72,1.00}
\definecolor{SteelBlue2}{rgb}{0.36,0.67,0.93}
\definecolor{SteelBlue3}{rgb}{0.31,0.58,0.80}
\definecolor{SteelBlue4}{rgb}{0.21,0.39,0.55}
\definecolor{SteelBlue}{rgb}{0.27,0.51,0.71}
\definecolor{VioletRed1}{rgb}{1.00,0.24,0.59}
\definecolor{VioletRed2}{rgb}{0.93,0.23,0.55}
\definecolor{VioletRed3}{rgb}{0.80,0.20,0.47}
\definecolor{VioletRed4}{rgb}{0.55,0.13,0.32}
\definecolor{VioletRed}{rgb}{0.82,0.13,0.56}
\definecolor{WhiteSmoke}{rgb}{0.96,0.96,0.96}
\definecolor{YellowGreen}{rgb}{0.60,0.80,0.20}
\definecolor{aliceblue}{rgb}{0.94,0.97,1.00}
\definecolor{antiquewhite}{rgb}{0.98,0.92,0.84}
\definecolor{aquamarine1}{rgb}{0.50,1.00,0.83}
\definecolor{aquamarine2}{rgb}{0.46,0.93,0.78}
\definecolor{aquamarine3}{rgb}{0.40,0.80,0.67}
\definecolor{aquamarine4}{rgb}{0.27,0.55,0.45}
\definecolor{aquamarine}{rgb}{0.50,1.00,0.83}
\definecolor{azure1}{rgb}{0.94,1.00,1.00}
\definecolor{azure2}{rgb}{0.88,0.93,0.93}
\definecolor{azure3}{rgb}{0.76,0.80,0.80}
\definecolor{azure4}{rgb}{0.51,0.55,0.55}
\definecolor{azure}{rgb}{0.94,1.00,1.00}
\definecolor{beige}{rgb}{0.96,0.96,0.86}
\definecolor{bisque1}{rgb}{1.00,0.89,0.77}
\definecolor{bisque2}{rgb}{0.93,0.84,0.72}
\definecolor{bisque3}{rgb}{0.80,0.72,0.62}
\definecolor{bisque4}{rgb}{0.55,0.49,0.42}
\definecolor{bisque}{rgb}{1.00,0.89,0.77}
\definecolor{black}{rgb}{0.00,0.00,0.00}
\definecolor{blanchedalmond}{rgb}{1.00,0.92,0.80}
\definecolor{blue1}{rgb}{0.00,0.00,1.00}
\definecolor{blue2}{rgb}{0.00,0.00,0.93}
\definecolor{blue3}{rgb}{0.00,0.00,0.80}
\definecolor{blue4}{rgb}{0.00,0.00,0.55}
\definecolor{blueviolet}{rgb}{0.54,0.17,0.89}
\definecolor{blue}{rgb}{0.00,0.00,1.00}
\definecolor{brown1}{rgb}{1.00,0.25,0.25}
\definecolor{brown2}{rgb}{0.93,0.23,0.23}
\definecolor{brown3}{rgb}{0.80,0.20,0.20}
\definecolor{brown4}{rgb}{0.55,0.14,0.14}
\definecolor{brown}{rgb}{0.65,0.16,0.16}
\definecolor{burlywood1}{rgb}{1.00,0.83,0.61}
\definecolor{burlywood2}{rgb}{0.93,0.77,0.57}
\definecolor{burlywood3}{rgb}{0.80,0.67,0.49}
\definecolor{burlywood4}{rgb}{0.55,0.45,0.33}
\definecolor{burlywood}{rgb}{0.87,0.72,0.53}
\definecolor{cadetblue}{rgb}{0.37,0.62,0.63}
\definecolor{chartreuse1}{rgb}{0.50,1.00,0.00}
\definecolor{chartreuse2}{rgb}{0.46,0.93,0.00}
\definecolor{chartreuse3}{rgb}{0.40,0.80,0.00}
\definecolor{chartreuse4}{rgb}{0.27,0.55,0.00}
\definecolor{chartreuse}{rgb}{0.50,1.00,0.00}
\definecolor{chocolate1}{rgb}{1.00,0.50,0.14}
\definecolor{chocolate2}{rgb}{0.93,0.46,0.13}
\definecolor{chocolate3}{rgb}{0.80,0.40,0.11}
\definecolor{chocolate4}{rgb}{0.55,0.27,0.07}
\definecolor{chocolate}{rgb}{0.82,0.41,0.12}
\definecolor{coral1}{rgb}{1.00,0.45,0.34}
\definecolor{coral2}{rgb}{0.93,0.42,0.31}
\definecolor{coral3}{rgb}{0.80,0.36,0.27}
\definecolor{coral4}{rgb}{0.55,0.24,0.18}
\definecolor{coral}{rgb}{1.00,0.50,0.31}
\definecolor{cornflowerblue}{rgb}{0.39,0.58,0.93}
\definecolor{cornsilk1}{rgb}{1.00,0.97,0.86}
\definecolor{cornsilk2}{rgb}{0.93,0.91,0.80}
\definecolor{cornsilk3}{rgb}{0.80,0.78,0.69}
\definecolor{cornsilk4}{rgb}{0.55,0.53,0.47}
\definecolor{cornsilk}{rgb}{1.00,0.97,0.86}
\definecolor{cyan1}{rgb}{0.00,1.00,1.00}
\definecolor{cyan2}{rgb}{0.00,0.93,0.93}
\definecolor{cyan3}{rgb}{0.00,0.80,0.80}
\definecolor{cyan4}{rgb}{0.00,0.55,0.55}
\definecolor{cyan}{rgb}{0.00,1.00,1.00}
\definecolor{darkblue}{rgb}{0.00,0.00,0.55}
\definecolor{darkcyan}{rgb}{0.00,0.55,0.55}
\definecolor{darkgoldenrod}{rgb}{0.72,0.53,0.04}
\definecolor{darkgray}{rgb}{0.66,0.66,0.66}
\definecolor{darkgreen}{rgb}{0.00,0.39,0.00}
\definecolor{darkgrey}{rgb}{0.66,0.66,0.66}
\definecolor{darkkhaki}{rgb}{0.74,0.72,0.42}
\definecolor{darkmagenta}{rgb}{0.55,0.00,0.55}
\definecolor{darkolive}{rgb}{0.33,0.42,0.18}
\definecolor{darkorange}{rgb}{1.00,0.55,0.00}
\definecolor{darkorchid}{rgb}{0.60,0.20,0.80}
\definecolor{darkred}{rgb}{0.55,0.00,0.00}
\definecolor{darksalmon}{rgb}{0.91,0.59,0.48}
\definecolor{darksea}{rgb}{0.56,0.74,0.56}
\definecolor{darkslate}{rgb}{0.18,0.31,0.31}
\definecolor{darkslate}{rgb}{0.18,0.31,0.31}
\definecolor{darkslate}{rgb}{0.28,0.24,0.55}
\definecolor{darkturquoise}{rgb}{0.00,0.81,0.82}
\definecolor{darkviolet}{rgb}{0.58,0.00,0.83}
\definecolor{deeppink}{rgb}{1.00,0.08,0.58}
\definecolor{deepsky}{rgb}{0.00,0.75,1.00}
\definecolor{dimgray}{rgb}{0.41,0.41,0.41}
\definecolor{dimgrey}{rgb}{0.41,0.41,0.41}
\definecolor{dodgerblue}{rgb}{0.12,0.56,1.00}
\definecolor{firebrick1}{rgb}{1.00,0.19,0.19}
\definecolor{firebrick2}{rgb}{0.93,0.17,0.17}
\definecolor{firebrick3}{rgb}{0.80,0.15,0.15}
\definecolor{firebrick4}{rgb}{0.55,0.10,0.10}
\definecolor{firebrick}{rgb}{0.70,0.13,0.13}
\definecolor{floralwhite}{rgb}{1.00,0.98,0.94}
\definecolor{forestgreen}{rgb}{0.13,0.55,0.13}
\definecolor{gainsboro}{rgb}{0.86,0.86,0.86}
\definecolor{ghostwhite}{rgb}{0.97,0.97,1.00}
\definecolor{gold1}{rgb}{1.00,0.84,0.00}
\definecolor{gold2}{rgb}{0.93,0.79,0.00}
\definecolor{gold3}{rgb}{0.80,0.68,0.00}
\definecolor{gold4}{rgb}{0.55,0.46,0.00}
\definecolor{goldenrod1}{rgb}{1.00,0.76,0.15}
\definecolor{goldenrod2}{rgb}{0.93,0.71,0.13}
\definecolor{goldenrod3}{rgb}{0.80,0.61,0.11}
\definecolor{goldenrod4}{rgb}{0.55,0.41,0.08}
\definecolor{goldenrod}{rgb}{0.85,0.65,0.13}
\definecolor{gold}{rgb}{1.00,0.84,0.00}
\definecolor{gray0}{rgb}{0.00,0.00,0.00}
\definecolor{gray100}{rgb}{1.00,1.00,1.00}
\definecolor{gray10}{rgb}{0.10,0.10,0.10}
\definecolor{gray11}{rgb}{0.11,0.11,0.11}
\definecolor{gray12}{rgb}{0.12,0.12,0.12}
\definecolor{gray13}{rgb}{0.13,0.13,0.13}
\definecolor{gray14}{rgb}{0.14,0.14,0.14}
\definecolor{gray15}{rgb}{0.15,0.15,0.15}
\definecolor{gray16}{rgb}{0.16,0.16,0.16}
\definecolor{gray17}{rgb}{0.17,0.17,0.17}
\definecolor{gray18}{rgb}{0.18,0.18,0.18}
\definecolor{gray19}{rgb}{0.19,0.19,0.19}
\definecolor{gray1}{rgb}{0.01,0.01,0.01}
\definecolor{gray20}{rgb}{0.20,0.20,0.20}
\definecolor{gray21}{rgb}{0.21,0.21,0.21}
\definecolor{gray22}{rgb}{0.22,0.22,0.22}
\definecolor{gray23}{rgb}{0.23,0.23,0.23}
\definecolor{gray24}{rgb}{0.24,0.24,0.24}
\definecolor{gray25}{rgb}{0.25,0.25,0.25}
\definecolor{gray26}{rgb}{0.26,0.26,0.26}
\definecolor{gray27}{rgb}{0.27,0.27,0.27}
\definecolor{gray28}{rgb}{0.28,0.28,0.28}
\definecolor{gray29}{rgb}{0.29,0.29,0.29}
\definecolor{gray2}{rgb}{0.02,0.02,0.02}
\definecolor{gray30}{rgb}{0.30,0.30,0.30}
\definecolor{gray31}{rgb}{0.31,0.31,0.31}
\definecolor{gray32}{rgb}{0.32,0.32,0.32}
\definecolor{gray33}{rgb}{0.33,0.33,0.33}
\definecolor{gray34}{rgb}{0.34,0.34,0.34}
\definecolor{gray35}{rgb}{0.35,0.35,0.35}
\definecolor{gray36}{rgb}{0.36,0.36,0.36}
\definecolor{gray37}{rgb}{0.37,0.37,0.37}
\definecolor{gray38}{rgb}{0.38,0.38,0.38}
\definecolor{gray39}{rgb}{0.39,0.39,0.39}
\definecolor{gray3}{rgb}{0.03,0.03,0.03}
\definecolor{gray40}{rgb}{0.40,0.40,0.40}
\definecolor{gray41}{rgb}{0.41,0.41,0.41}
\definecolor{gray42}{rgb}{0.42,0.42,0.42}
\definecolor{gray43}{rgb}{0.43,0.43,0.43}
\definecolor{gray44}{rgb}{0.44,0.44,0.44}
\definecolor{gray45}{rgb}{0.45,0.45,0.45}
\definecolor{gray46}{rgb}{0.46,0.46,0.46}
\definecolor{gray47}{rgb}{0.47,0.47,0.47}
\definecolor{gray48}{rgb}{0.48,0.48,0.48}
\definecolor{gray49}{rgb}{0.49,0.49,0.49}
\definecolor{gray4}{rgb}{0.04,0.04,0.04}
\definecolor{gray50}{rgb}{0.50,0.50,0.50}
\definecolor{gray51}{rgb}{0.51,0.51,0.51}
\definecolor{gray52}{rgb}{0.52,0.52,0.52}
\definecolor{gray53}{rgb}{0.53,0.53,0.53}
\definecolor{gray54}{rgb}{0.54,0.54,0.54}
\definecolor{gray55}{rgb}{0.55,0.55,0.55}
\definecolor{gray56}{rgb}{0.56,0.56,0.56}
\definecolor{gray57}{rgb}{0.57,0.57,0.57}
\definecolor{gray58}{rgb}{0.58,0.58,0.58}
\definecolor{gray59}{rgb}{0.59,0.59,0.59}
\definecolor{gray5}{rgb}{0.05,0.05,0.05}
\definecolor{gray60}{rgb}{0.60,0.60,0.60}
\definecolor{gray61}{rgb}{0.61,0.61,0.61}
\definecolor{gray62}{rgb}{0.62,0.62,0.62}
\definecolor{gray63}{rgb}{0.63,0.63,0.63}
\definecolor{gray64}{rgb}{0.64,0.64,0.64}
\definecolor{gray65}{rgb}{0.65,0.65,0.65}
\definecolor{gray66}{rgb}{0.66,0.66,0.66}
\definecolor{gray67}{rgb}{0.67,0.67,0.67}
\definecolor{gray68}{rgb}{0.68,0.68,0.68}
\definecolor{gray69}{rgb}{0.69,0.69,0.69}
\definecolor{gray6}{rgb}{0.06,0.06,0.06}
\definecolor{gray70}{rgb}{0.70,0.70,0.70}
\definecolor{gray71}{rgb}{0.71,0.71,0.71}
\definecolor{gray72}{rgb}{0.72,0.72,0.72}
\definecolor{gray73}{rgb}{0.73,0.73,0.73}
\definecolor{gray74}{rgb}{0.74,0.74,0.74}
\definecolor{gray75}{rgb}{0.75,0.75,0.75}
\definecolor{gray76}{rgb}{0.76,0.76,0.76}
\definecolor{gray77}{rgb}{0.77,0.77,0.77}
\definecolor{gray78}{rgb}{0.78,0.78,0.78}
\definecolor{gray79}{rgb}{0.79,0.79,0.79}
\definecolor{gray7}{rgb}{0.07,0.07,0.07}
\definecolor{gray80}{rgb}{0.80,0.80,0.80}
\definecolor{gray81}{rgb}{0.81,0.81,0.81}
\definecolor{gray82}{rgb}{0.82,0.82,0.82}
\definecolor{gray83}{rgb}{0.83,0.83,0.83}
\definecolor{gray84}{rgb}{0.84,0.84,0.84}
\definecolor{gray85}{rgb}{0.85,0.85,0.85}
\definecolor{gray86}{rgb}{0.86,0.86,0.86}
\definecolor{gray87}{rgb}{0.87,0.87,0.87}
\definecolor{gray88}{rgb}{0.88,0.88,0.88}
\definecolor{gray89}{rgb}{0.89,0.89,0.89}
\definecolor{gray8}{rgb}{0.08,0.08,0.08}
\definecolor{gray90}{rgb}{0.90,0.90,0.90}
\definecolor{gray91}{rgb}{0.91,0.91,0.91}
\definecolor{gray92}{rgb}{0.92,0.92,0.92}
\definecolor{gray93}{rgb}{0.93,0.93,0.93}
\definecolor{gray94}{rgb}{0.94,0.94,0.94}
\definecolor{gray95}{rgb}{0.95,0.95,0.95}
\definecolor{gray96}{rgb}{0.96,0.96,0.96}
\definecolor{gray97}{rgb}{0.97,0.97,0.97}
\definecolor{gray98}{rgb}{0.98,0.98,0.98}
\definecolor{gray99}{rgb}{0.99,0.99,0.99}
\definecolor{gray9}{rgb}{0.09,0.09,0.09}
\definecolor{gray}{rgb}{0.75,0.75,0.75}
\definecolor{green1}{rgb}{0.00,1.00,0.00}
\definecolor{green2}{rgb}{0.00,0.93,0.00}
\definecolor{green3}{rgb}{0.00,0.80,0.00}
\definecolor{green4}{rgb}{0.00,0.55,0.00}
\definecolor{greenyellow}{rgb}{0.68,1.00,0.18}
\definecolor{green}{rgb}{0.00,1.00,0.00}
\definecolor{grey0}{rgb}{0.00,0.00,0.00}
\definecolor{grey100}{rgb}{1.00,1.00,1.00}
\definecolor{grey10}{rgb}{0.10,0.10,0.10}
\definecolor{grey11}{rgb}{0.11,0.11,0.11}
\definecolor{grey12}{rgb}{0.12,0.12,0.12}
\definecolor{grey13}{rgb}{0.13,0.13,0.13}
\definecolor{grey14}{rgb}{0.14,0.14,0.14}
\definecolor{grey15}{rgb}{0.15,0.15,0.15}
\definecolor{grey16}{rgb}{0.16,0.16,0.16}
\definecolor{grey17}{rgb}{0.17,0.17,0.17}
\definecolor{grey18}{rgb}{0.18,0.18,0.18}
\definecolor{grey19}{rgb}{0.19,0.19,0.19}
\definecolor{grey1}{rgb}{0.01,0.01,0.01}
\definecolor{grey20}{rgb}{0.20,0.20,0.20}
\definecolor{grey21}{rgb}{0.21,0.21,0.21}
\definecolor{grey22}{rgb}{0.22,0.22,0.22}
\definecolor{grey23}{rgb}{0.23,0.23,0.23}
\definecolor{grey24}{rgb}{0.24,0.24,0.24}
\definecolor{grey25}{rgb}{0.25,0.25,0.25}
\definecolor{grey26}{rgb}{0.26,0.26,0.26}
\definecolor{grey27}{rgb}{0.27,0.27,0.27}
\definecolor{grey28}{rgb}{0.28,0.28,0.28}
\definecolor{grey29}{rgb}{0.29,0.29,0.29}
\definecolor{grey2}{rgb}{0.02,0.02,0.02}
\definecolor{grey30}{rgb}{0.30,0.30,0.30}
\definecolor{grey31}{rgb}{0.31,0.31,0.31}
\definecolor{grey32}{rgb}{0.32,0.32,0.32}
\definecolor{grey33}{rgb}{0.33,0.33,0.33}
\definecolor{grey34}{rgb}{0.34,0.34,0.34}
\definecolor{grey35}{rgb}{0.35,0.35,0.35}
\definecolor{grey36}{rgb}{0.36,0.36,0.36}
\definecolor{grey37}{rgb}{0.37,0.37,0.37}
\definecolor{grey38}{rgb}{0.38,0.38,0.38}
\definecolor{grey39}{rgb}{0.39,0.39,0.39}
\definecolor{grey3}{rgb}{0.03,0.03,0.03}
\definecolor{grey40}{rgb}{0.40,0.40,0.40}
\definecolor{grey41}{rgb}{0.41,0.41,0.41}
\definecolor{grey42}{rgb}{0.42,0.42,0.42}
\definecolor{grey43}{rgb}{0.43,0.43,0.43}
\definecolor{grey44}{rgb}{0.44,0.44,0.44}
\definecolor{grey45}{rgb}{0.45,0.45,0.45}
\definecolor{grey46}{rgb}{0.46,0.46,0.46}
\definecolor{grey47}{rgb}{0.47,0.47,0.47}
\definecolor{grey48}{rgb}{0.48,0.48,0.48}
\definecolor{grey49}{rgb}{0.49,0.49,0.49}
\definecolor{grey4}{rgb}{0.04,0.04,0.04}
\definecolor{grey50}{rgb}{0.50,0.50,0.50}
\definecolor{grey51}{rgb}{0.51,0.51,0.51}
\definecolor{grey52}{rgb}{0.52,0.52,0.52}
\definecolor{grey53}{rgb}{0.53,0.53,0.53}
\definecolor{grey54}{rgb}{0.54,0.54,0.54}
\definecolor{grey55}{rgb}{0.55,0.55,0.55}
\definecolor{grey56}{rgb}{0.56,0.56,0.56}
\definecolor{grey57}{rgb}{0.57,0.57,0.57}
\definecolor{grey58}{rgb}{0.58,0.58,0.58}
\definecolor{grey59}{rgb}{0.59,0.59,0.59}
\definecolor{grey5}{rgb}{0.05,0.05,0.05}
\definecolor{grey60}{rgb}{0.60,0.60,0.60}
\definecolor{grey61}{rgb}{0.61,0.61,0.61}
\definecolor{grey62}{rgb}{0.62,0.62,0.62}
\definecolor{grey63}{rgb}{0.63,0.63,0.63}
\definecolor{grey64}{rgb}{0.64,0.64,0.64}
\definecolor{grey65}{rgb}{0.65,0.65,0.65}
\definecolor{grey66}{rgb}{0.66,0.66,0.66}
\definecolor{grey67}{rgb}{0.67,0.67,0.67}
\definecolor{grey68}{rgb}{0.68,0.68,0.68}
\definecolor{grey69}{rgb}{0.69,0.69,0.69}
\definecolor{grey6}{rgb}{0.06,0.06,0.06}
\definecolor{grey70}{rgb}{0.70,0.70,0.70}
\definecolor{grey71}{rgb}{0.71,0.71,0.71}
\definecolor{grey72}{rgb}{0.72,0.72,0.72}
\definecolor{grey73}{rgb}{0.73,0.73,0.73}
\definecolor{grey74}{rgb}{0.74,0.74,0.74}
\definecolor{grey75}{rgb}{0.75,0.75,0.75}
\definecolor{grey76}{rgb}{0.76,0.76,0.76}
\definecolor{grey77}{rgb}{0.77,0.77,0.77}
\definecolor{grey78}{rgb}{0.78,0.78,0.78}
\definecolor{grey79}{rgb}{0.79,0.79,0.79}
\definecolor{grey7}{rgb}{0.07,0.07,0.07}
\definecolor{grey80}{rgb}{0.80,0.80,0.80}
\definecolor{grey81}{rgb}{0.81,0.81,0.81}
\definecolor{grey82}{rgb}{0.82,0.82,0.82}
\definecolor{grey83}{rgb}{0.83,0.83,0.83}
\definecolor{grey84}{rgb}{0.84,0.84,0.84}
\definecolor{grey85}{rgb}{0.85,0.85,0.85}
\definecolor{grey86}{rgb}{0.86,0.86,0.86}
\definecolor{grey87}{rgb}{0.87,0.87,0.87}
\definecolor{grey88}{rgb}{0.88,0.88,0.88}
\definecolor{grey89}{rgb}{0.89,0.89,0.89}
\definecolor{grey8}{rgb}{0.08,0.08,0.08}
\definecolor{grey90}{rgb}{0.90,0.90,0.90}
\definecolor{grey91}{rgb}{0.91,0.91,0.91}
\definecolor{grey92}{rgb}{0.92,0.92,0.92}
\definecolor{grey93}{rgb}{0.93,0.93,0.93}
\definecolor{grey94}{rgb}{0.94,0.94,0.94}
\definecolor{grey95}{rgb}{0.95,0.95,0.95}
\definecolor{grey96}{rgb}{0.96,0.96,0.96}
\definecolor{grey97}{rgb}{0.97,0.97,0.97}
\definecolor{grey98}{rgb}{0.98,0.98,0.98}
\definecolor{grey99}{rgb}{0.99,0.99,0.99}
\definecolor{grey9}{rgb}{0.09,0.09,0.09}
\definecolor{grey}{rgb}{0.75,0.75,0.75}
\definecolor{honeydew1}{rgb}{0.94,1.00,0.94}
\definecolor{honeydew2}{rgb}{0.88,0.93,0.88}
\definecolor{honeydew3}{rgb}{0.76,0.80,0.76}
\definecolor{honeydew4}{rgb}{0.51,0.55,0.51}
\definecolor{honeydew}{rgb}{0.94,1.00,0.94}
\definecolor{hotpink}{rgb}{1.00,0.41,0.71}
\definecolor{indianred}{rgb}{0.80,0.36,0.36}
\definecolor{ivory1}{rgb}{1.00,1.00,0.94}
\definecolor{ivory2}{rgb}{0.93,0.93,0.88}
\definecolor{ivory3}{rgb}{0.80,0.80,0.76}
\definecolor{ivory4}{rgb}{0.55,0.55,0.51}
\definecolor{ivory}{rgb}{1.00,1.00,0.94}
\definecolor{khaki1}{rgb}{1.00,0.96,0.56}
\definecolor{khaki2}{rgb}{0.93,0.90,0.52}
\definecolor{khaki3}{rgb}{0.80,0.78,0.45}
\definecolor{khaki4}{rgb}{0.55,0.53,0.31}
\definecolor{khaki}{rgb}{0.94,0.90,0.55}
\definecolor{lavenderblush}{rgb}{1.00,0.94,0.96}
\definecolor{lavender}{rgb}{0.90,0.90,0.98}
\definecolor{lawngreen}{rgb}{0.49,0.99,0.00}
\definecolor{lemonchiffon}{rgb}{1.00,0.98,0.80}
\definecolor{lightblue}{rgb}{0.68,0.85,0.90}
\definecolor{lightcoral}{rgb}{0.94,0.50,0.50}
\definecolor{lightcyan}{rgb}{0.88,1.00,1.00}
\definecolor{lightgoldenrod}{rgb}{0.93,0.87,0.51}
\definecolor{lightgoldenrod}{rgb}{0.98,0.98,0.82}
\definecolor{lightgray}{rgb}{0.83,0.83,0.83}
\definecolor{lightgreen}{rgb}{0.56,0.93,0.56}
\definecolor{lightgrey}{rgb}{0.83,0.83,0.83}
\definecolor{lightpink}{rgb}{1.00,0.71,0.76}
\definecolor{lightsalmon}{rgb}{1.00,0.63,0.48}
\definecolor{lightsea}{rgb}{0.13,0.70,0.67}
\definecolor{lightsky}{rgb}{0.53,0.81,0.98}
\definecolor{lightslate}{rgb}{0.47,0.53,0.60}
\definecolor{lightslate}{rgb}{0.47,0.53,0.60}
\definecolor{lightslate}{rgb}{0.52,0.44,1.00}
\definecolor{lightsteel}{rgb}{0.69,0.77,0.87}
\definecolor{lightyellow}{rgb}{1.00,1.00,0.88}
\definecolor{limegreen}{rgb}{0.20,0.80,0.20}
\definecolor{linen}{rgb}{0.98,0.94,0.90}
\definecolor{magenta1}{rgb}{1.00,0.00,1.00}
\definecolor{magenta2}{rgb}{0.93,0.00,0.93}
\definecolor{magenta3}{rgb}{0.80,0.00,0.80}
\definecolor{magenta4}{rgb}{0.55,0.00,0.55}
\definecolor{magenta}{rgb}{1.00,0.00,1.00}
\definecolor{maroon1}{rgb}{1.00,0.20,0.70}
\definecolor{maroon2}{rgb}{0.93,0.19,0.65}
\definecolor{maroon3}{rgb}{0.80,0.16,0.56}
\definecolor{maroon4}{rgb}{0.55,0.11,0.38}
\definecolor{maroon}{rgb}{0.69,0.19,0.38}
\definecolor{mediumaquamarine}{rgb}{0.40,0.80,0.67}
\definecolor{mediumblue}{rgb}{0.00,0.00,0.80}
\definecolor{mediumorchid}{rgb}{0.73,0.33,0.83}
\definecolor{mediumpurple}{rgb}{0.58,0.44,0.86}
\definecolor{mediumsea}{rgb}{0.24,0.70,0.44}
\definecolor{mediumslate}{rgb}{0.48,0.41,0.93}
\definecolor{mediumspring}{rgb}{0.00,0.98,0.60}
\definecolor{mediumturquoise}{rgb}{0.28,0.82,0.80}
\definecolor{mediumviolet}{rgb}{0.78,0.08,0.52}
\definecolor{midnightblue}{rgb}{0.10,0.10,0.44}
\definecolor{mintcream}{rgb}{0.96,1.00,0.98}
\definecolor{mistyrose}{rgb}{1.00,0.89,0.88}
\definecolor{moccasin}{rgb}{1.00,0.89,0.71}
\definecolor{navajowhite}{rgb}{1.00,0.87,0.68}
\definecolor{navyblue}{rgb}{0.00,0.00,0.50}
\definecolor{navy}{rgb}{0.00,0.00,0.50}
\definecolor{oldlace}{rgb}{0.99,0.96,0.90}
\definecolor{olivedrab}{rgb}{0.42,0.56,0.14}
\definecolor{orange1}{rgb}{1.00,0.65,0.00}
\definecolor{orange2}{rgb}{0.93,0.60,0.00}
\definecolor{orange3}{rgb}{0.80,0.52,0.00}
\definecolor{orange4}{rgb}{0.55,0.35,0.00}
\definecolor{orangered}{rgb}{1.00,0.27,0.00}
\definecolor{orange}{rgb}{1.00,0.65,0.00}
\definecolor{orchid1}{rgb}{1.00,0.51,0.98}
\definecolor{orchid2}{rgb}{0.93,0.48,0.91}
\definecolor{orchid3}{rgb}{0.80,0.41,0.79}
\definecolor{orchid4}{rgb}{0.55,0.28,0.54}
\definecolor{orchid}{rgb}{0.85,0.44,0.84}
\definecolor{palegoldenrod}{rgb}{0.93,0.91,0.67}
\definecolor{palegreen}{rgb}{0.60,0.98,0.60}
\definecolor{paleturquoise}{rgb}{0.69,0.93,0.93}
\definecolor{paleviolet}{rgb}{0.86,0.44,0.58}
\definecolor{papayawhip}{rgb}{1.00,0.94,0.84}
\definecolor{peachpuff}{rgb}{1.00,0.85,0.73}
\definecolor{peru}{rgb}{0.80,0.52,0.25}
\definecolor{pink1}{rgb}{1.00,0.71,0.77}
\definecolor{pink2}{rgb}{0.93,0.66,0.72}
\definecolor{pink3}{rgb}{0.80,0.57,0.62}
\definecolor{pink4}{rgb}{0.55,0.39,0.42}
\definecolor{pink}{rgb}{1.00,0.75,0.80}
\definecolor{plum1}{rgb}{1.00,0.73,1.00}
\definecolor{plum2}{rgb}{0.93,0.68,0.93}
\definecolor{plum3}{rgb}{0.80,0.59,0.80}
\definecolor{plum4}{rgb}{0.55,0.40,0.55}
\definecolor{plum}{rgb}{0.87,0.63,0.87}
\definecolor{powderblue}{rgb}{0.69,0.88,0.90}
\definecolor{purple1}{rgb}{0.61,0.19,1.00}
\definecolor{purple2}{rgb}{0.57,0.17,0.93}
\definecolor{purple3}{rgb}{0.49,0.15,0.80}
\definecolor{purple4}{rgb}{0.33,0.10,0.55}
\definecolor{purple}{rgb}{0.63,0.13,0.94}
\definecolor{red1}{rgb}{1.00,0.00,0.00}
\definecolor{red2}{rgb}{0.93,0.00,0.00}
\definecolor{red3}{rgb}{0.80,0.00,0.00}
\definecolor{red4}{rgb}{0.55,0.00,0.00}
\definecolor{red}{rgb}{1.00,0.00,0.00}
\definecolor{rosybrown}{rgb}{0.74,0.56,0.56}
\definecolor{royalblue}{rgb}{0.25,0.41,0.88}
\definecolor{saddlebrown}{rgb}{0.55,0.27,0.07}
\definecolor{salmon1}{rgb}{1.00,0.55,0.41}
\definecolor{salmon2}{rgb}{0.93,0.51,0.38}
\definecolor{salmon3}{rgb}{0.80,0.44,0.33}
\definecolor{salmon4}{rgb}{0.55,0.30,0.22}
\definecolor{salmon}{rgb}{0.98,0.50,0.45}
\definecolor{sandybrown}{rgb}{0.96,0.64,0.38}
\definecolor{seagreen}{rgb}{0.18,0.55,0.34}
\definecolor{seashell1}{rgb}{1.00,0.96,0.93}
\definecolor{seashell2}{rgb}{0.93,0.90,0.87}
\definecolor{seashell3}{rgb}{0.80,0.77,0.75}
\definecolor{seashell4}{rgb}{0.55,0.53,0.51}
\definecolor{seashell}{rgb}{1.00,0.96,0.93}
\definecolor{sienna1}{rgb}{1.00,0.51,0.28}
\definecolor{sienna2}{rgb}{0.93,0.47,0.26}
\definecolor{sienna3}{rgb}{0.80,0.41,0.22}
\definecolor{sienna4}{rgb}{0.55,0.28,0.15}
\definecolor{sienna}{rgb}{0.63,0.32,0.18}
\definecolor{skyblue}{rgb}{0.53,0.81,0.92}
\definecolor{slateblue}{rgb}{0.42,0.35,0.80}
\definecolor{slategray}{rgb}{0.44,0.50,0.56}
\definecolor{slategrey}{rgb}{0.44,0.50,0.56}
\definecolor{snow1}{rgb}{1.00,0.98,0.98}
\definecolor{snow2}{rgb}{0.93,0.91,0.91}
\definecolor{snow3}{rgb}{0.80,0.79,0.79}
\definecolor{snow4}{rgb}{0.55,0.54,0.54}
\definecolor{snow}{rgb}{1.00,0.98,0.98}
\definecolor{springgreen}{rgb}{0.00,1.00,0.50}
\definecolor{steelblue}{rgb}{0.27,0.51,0.71}
\definecolor{tan1}{rgb}{1.00,0.65,0.31}
\definecolor{tan2}{rgb}{0.93,0.60,0.29}
\definecolor{tan3}{rgb}{0.80,0.52,0.25}
\definecolor{tan4}{rgb}{0.55,0.35,0.17}
\definecolor{tan}{rgb}{0.82,0.71,0.55}
\definecolor{thistle1}{rgb}{1.00,0.88,1.00}
\definecolor{thistle2}{rgb}{0.93,0.82,0.93}
\definecolor{thistle3}{rgb}{0.80,0.71,0.80}
\definecolor{thistle4}{rgb}{0.55,0.48,0.55}
\definecolor{thistle}{rgb}{0.85,0.75,0.85}
\definecolor{tomato1}{rgb}{1.00,0.39,0.28}
\definecolor{tomato2}{rgb}{0.93,0.36,0.26}
\definecolor{tomato3}{rgb}{0.80,0.31,0.22}
\definecolor{tomato4}{rgb}{0.55,0.21,0.15}
\definecolor{tomato}{rgb}{1.00,0.39,0.28}
\definecolor{turquoise1}{rgb}{0.00,0.96,1.00}
\definecolor{turquoise2}{rgb}{0.00,0.90,0.93}
\definecolor{turquoise3}{rgb}{0.00,0.77,0.80}
\definecolor{turquoise4}{rgb}{0.00,0.53,0.55}
\definecolor{turquoise}{rgb}{0.25,0.88,0.82}
\definecolor{violetred}{rgb}{0.82,0.13,0.56}
\definecolor{violet}{rgb}{0.93,0.51,0.93}
\definecolor{wheat1}{rgb}{1.00,0.91,0.73}
\definecolor{wheat2}{rgb}{0.93,0.85,0.68}
\definecolor{wheat3}{rgb}{0.80,0.73,0.59}
\definecolor{wheat4}{rgb}{0.55,0.49,0.40}
\definecolor{wheat}{rgb}{0.96,0.87,0.70}
\definecolor{whitesmoke}{rgb}{0.96,0.96,0.96}
\definecolor{white}{rgb}{1.00,1.00,1.00}
\definecolor{yellow1}{rgb}{1.00,1.00,0.00}
\definecolor{yellow2}{rgb}{0.93,0.93,0.00}
\definecolor{yellow3}{rgb}{0.80,0.80,0.00}
\definecolor{yellow4}{rgb}{0.55,0.55,0.00}
\definecolor{yellowgreen}{rgb}{0.60,0.80,0.20}
\definecolor{yellow}{rgb}{1.00,1.00,0.00}

\title[Fisher Matrix Decomposition]{Fisher Matrix Decomposition for Dark Energy Prediction}

\author[Kitching \& Amara]{T. D. Kitching$^{1}$\thanks{tdk@roe.ac.uk}, A. Amara$^{2}$\\
$^{1}$SUPA\thanks{Scottish Universities Physics Alliance}, Institute for Astronomy, University of Edinburgh, Royal Observatory Edinburgh, Blackford Hill, EH9 3HJ\\
$^{2}$Department of Physics, ETH Zurich, Wolfgang-Pauli-Strasse 16,
  CH-8093 Zurich, Switzerland\\
}

\newcommand{\be}{\begin{equation}}
\newcommand{\ee}{\end{equation}}
\newcommand{\ba}{\begin{eqnarray}}
\newcommand{\ea}{\end{eqnarray}}
\newcommand{\nn}{\nonumber \\}
\newcommand{\ic}{{\tt iCosmo }}

\def\gs{\mathrel{\raise1.16pt\hbox{$>$}\kern-7.0pt %
\lower3.06pt\hbox{{$\scriptstyle \sim$}}}}         %
\def\ls{\mathrel{\raise1.16pt\hbox{$<$}\kern-7.0pt %
\lower3.06pt\hbox{{$\scriptstyle \sim$}}}}         %

\newcommand{\phib}{\mbox{\boldmath $\phi$}}

\newcommand{\psib}{\mbox{\boldmath $\psi$}}
\newcommand{\bw}{\mbox{\boldmath $w$}}

\def\gs{\mathrel{\raise1.16pt\hbox{$>$}\kern-7.0pt %
\lower3.06pt\hbox{{$\scriptstyle \sim$}}}}         %
\def\ls{\mathrel{\raise1.16pt\hbox{$<$}\kern-7.0pt %
\lower3.06pt\hbox{{$\scriptstyle \sim$}}}}         %

\begin{document}

\maketitle

\begin{abstract}
Within the context of constraining an expansion of the dark energy
equation of state $w(z)$ we show that the eigendecomposition of
Fisher matrices is 
sensitive to both the maximum order of the expansion and the
basis set choice. 
We investigate the Fisher matrix formalism in the case that a
particular function is expanded in some basis set. As an example we 
show results for an all sky weak lensing tomographic experiment.
We show that the set of eigenfunctions is not unique and
that the best constrained functions are only reproduced
accurately at very higher order $N \gs 100$, a tophat basis set
requires an even higher order. We show that the
common approach used for finding the marginalised eigenfunction
errors is sensitive to the choice of non-$w(z)$ parameters and priors. 
The eigendecomposition of Fisher matrices is a potentially useful
tool that can be used to determine the predicted accuracy with
which an experiment could constrain $w(z)$. It also allows for
the reconstruction of the redshift sensitivity of
the experiment to changes in $w(z)$. However the technique is
sensitive to both the order and the basis set choice. Publicly
available code is available as part of \ic at 
{\tt http://www.icosmo.org}.
\end{abstract}


\section{Introduction}
\label{Introduction}

The Fisher matrix methodology (Tegmark, Taylor \& Heavens, 1997;
Jungman et al., 1996; Fisher, 1935) is a statistical tool that has 
been used with some success in predicting the ability of future
experiments to constrain particular parameters of interest. In
cosmology Fisher matrices have gained some importance in predicting
the potential outcome of experiments, in particular dark energy surveys,
on which a large amount of resources may be spent. It is therefore of
paramount importance that the way in which Fisher matrices are used
should be understood and that any results that depend on this
methodology should be robust and reliable. 

In this article we will outline how the techniques of decomposing a general
matrix have a special interpretation when used within the Fisher
matrix framework. This will be done within the context of attempting
to predict constraints for  an extended parameter set. We will use the specific
example of attempting to constrain dark energy equation of state
$w(z)$ parameters (Albrecht \& Bernstein, 2007;
Tang et al., 2008; Simpson \& Bridle, 2006; 
Crittenden \& Pogosian, 2005; de Putter \& Linder, 2008; Dick et al., 2006; 
Huterer \& Cooray, 2005; Huterer \& Starkman, 2003; Hu, 2002; 
Knox et al., 2005; Ishak et al. 2006; Zhan et al., 2009a; Zhan et al.,
2009b; Cunha et al., 2009; Joudaki et al., 2009; Sarkar et al.,
2008), most recently Albrecht et al. (2009) defined a new dark energy
Figure of Merit using a binned $36$ parameter model.
 
We will show that the eigenfunctions obtained
by diagonalising a Fisher matrix are dependent of the basis set used in the
construction of the Fisher matrix, and that the eigenfunctions only
tend to agreement in the limit of a very large order of
expansion. We will also show that the usual method of marginalising
over extra parameters -- constructing the Schur complement (Zhang,
2005) of the
total Fisher matrix -- is dependent on the choice of non-$w(z)$
parameterisation and priors when also
performing an eigendecomposition.

In Section \ref{Methodology} 
we will introduce the methodology, in Section \ref{Application} we
will apply the approach to weak lensing tomographic survey prediction,
we present conclusions in Section \ref{Conclusion}.

\section{Methodology}
\label{Methodology}

For a set of parameters $\btheta$ the Fisher matrix allows for 
the prediction of parameter errors given a specific experimental design and 
method for extracting parameters. In the case of Gaussian-distributed
data where we assume that the error on the signal is not a function of 
parameter values 
$\sigma_C\not=\sigma_C(\btheta)$ we can take the covariance 
of the estimated values of the parameters; 
\be 
{\rm cov}[\btheta_i,\btheta_j]=\langle (\btheta_i-\langle\btheta_i\rangle)
(\btheta_j-\langle\btheta_j\rangle)\rangle=F^{-1}_{ij},
\ee
where the Fisher matrix is defined by 
(Tegmark, Taylor \& Heavens, 1997; Jungman et al., 1996; Fisher, 1935)
\be
\label{fishe}
F_{ij}=\sum_x \left[\sigma^{-2}_C\frac{\partial S}{\partial \btheta_i}
\frac{\partial S}{\partial \btheta_j}\right],
\ee
where $S$ is the signal. 
The predicted marginal errors on 
the parameters are given by $\Delta\theta_i=\sqrt{(F^{-1})_{ii}}$, 
this is the minimum marginal error that one can expect for the method
and experimental 
design considered (due to the lower bound in the Cramer-Rao inequality).  

If we extend the parameter set to $\phib=(\btheta$, $\psib)$ then the Fisher
matrix is extended to include these extra parameters
\be
\label{fishf}
F^{\phib\phib}=
\left( \begin{array}{cc}
 F^{\btheta\btheta} & F^{\btheta\psib}  \\
F^{\psib\btheta}   & F^{\psib\psib} \\
  \end{array}\right)
\ee
and the predicted marginal errors on the parameters 
$\Delta\theta_i=\sqrt{(F^{-1})^{\phib\phib}_{ii}}$ are increased due
to degeneracies between the original parameters and the new
parameters. 

For dark energy cosmology prediction the parameter set is commonly
divided into those that contain information on the dark energy
equation of state $w(z)$ and those that do not. The usual approach
taken
(Albrecht et al., 2009; Albrecht \& Bernstein, 2007;
Tang et al., 2008; Simpson \& Bridle, 2006; 
Crittenden \& Pogosian, 2005; de Putter \& Linder, 2008; Dick et al., 2006; 
Huterer \& Cooray, 2005; Huterer \& Starkman, 2003; Hu, 2002; 
Knox et al., 2005; Ishak et al. 2006; 
Zhan et al., 2009a; Zhan et al., 2009b; Cunha et al., 2009; Joudaki et
al., 2009; Sarkar et al., 2008)
is to 
``bin'' $w(z)$ in redshift and to assign an amplitude to
the value of $w(z)$ in each bin. However $w(z)$ is not
a directly observable quantity but can only be inferred through other
integral relations, most simply the comoving distance or Hubble parameter
\be 
D(z)=\frac{c}{H_0}\int_0^z {\rm d}z
\frac{1}{(\Omega_m(1+z)^3+\Omega_{\rm de}{\rm e}^{3 \int_0^z {\rm d}z'
    [1+w(z')]/(1+z')})^{\frac{1}{2}}}
\ee
\be
H(z)=H_0(\Omega_m(1+z)^3+\Omega_{\rm de}{\rm e}^{3 \int_0^z {\rm d}z'
    [1+w(z')]/(1+z')})^{\frac{1}{2}}.
\ee 
So any binned expansion of $w(z)$ to include extra
parameters cannot be 
justified by arguing that the data itself is binned in
redshift.

The choice of how to expand the function $w(z)$ is thus a purely
theoretical one. In general 
we will consider $w(z)$ expanded in some complete basis set 
\be 
\label{expand}
1+w(z)=\sum_{i=1}^{\infty}a_n\phi_n(z),
\ee
where $a_n$ are the basis coefficients and $\phi_n$ are the basis
functions. Since we do not know what dark energy equation state is apt
for the physical Universe any chosen basis set is as valid as any
other. Binning is a special case where the basis functions are tophat
functions in redshift 
\be 
\phi_n^{\rm binning}(z)={\mathcal H}(z-\frac{\Delta z}{2})-{\mathcal H}(z+\frac{\Delta z}{2}),
\ee
where $\Delta z$ is the bin width and ${\mathcal H}$ is the Heaviside (step)
function.

\begin{figure*}
\begin{center}
\resizebox{54mm}{!}{\psfig{file=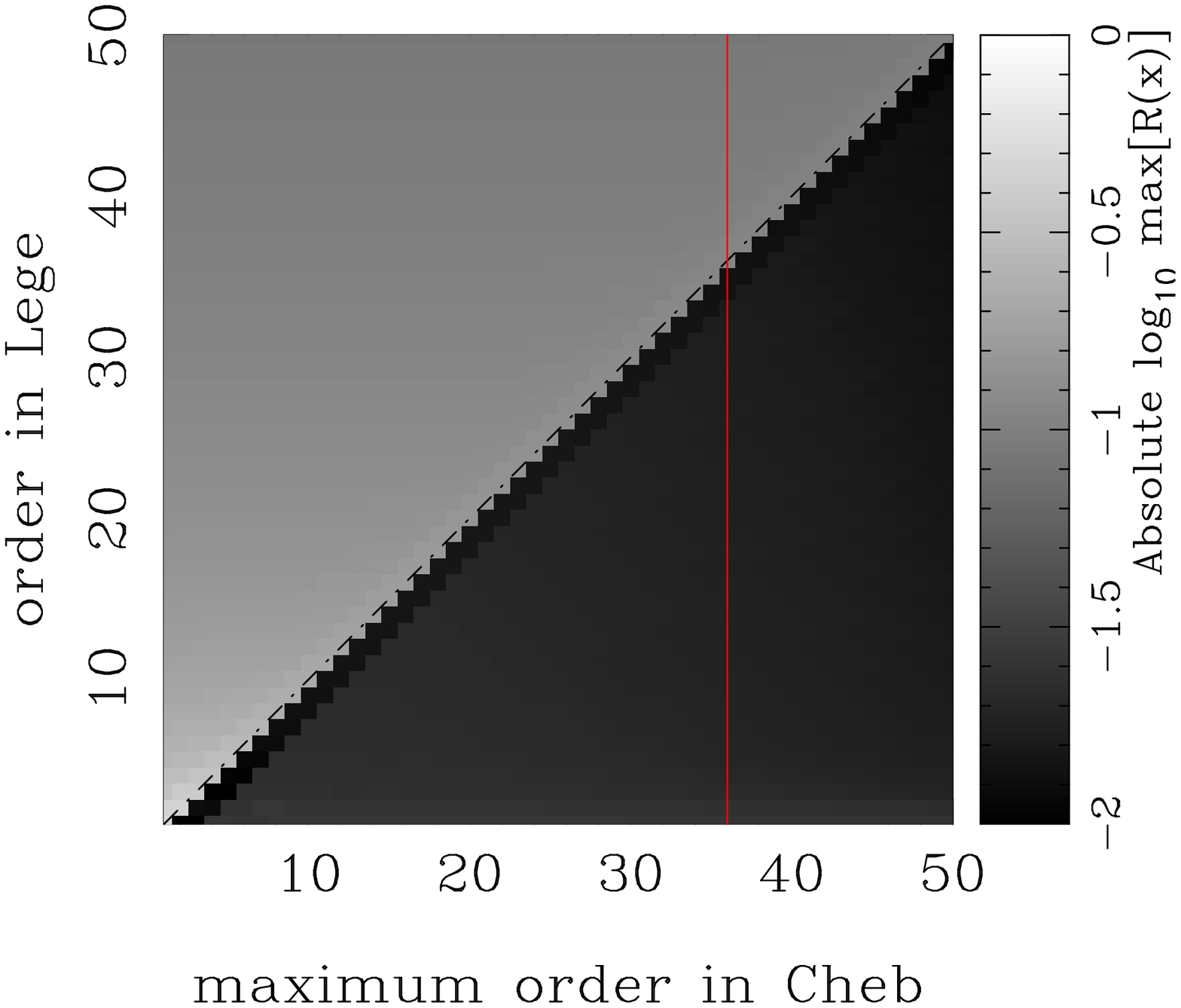, width=0.333\columnwidth, angle=0}}
\resizebox{54mm}{!}{\psfig{file=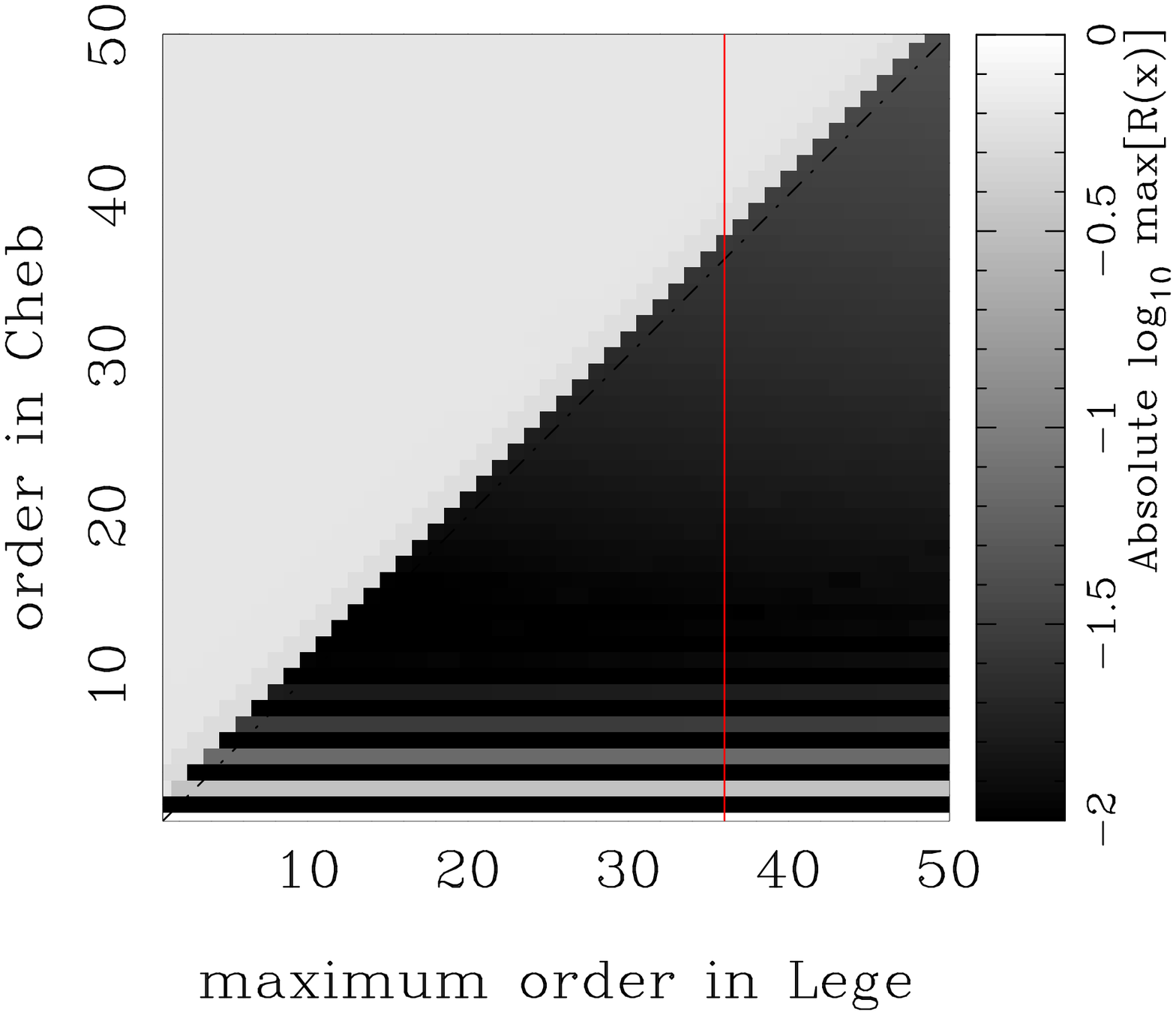, width=0.333\columnwidth, angle=0}}
\resizebox{54mm}{!}{\psfig{file=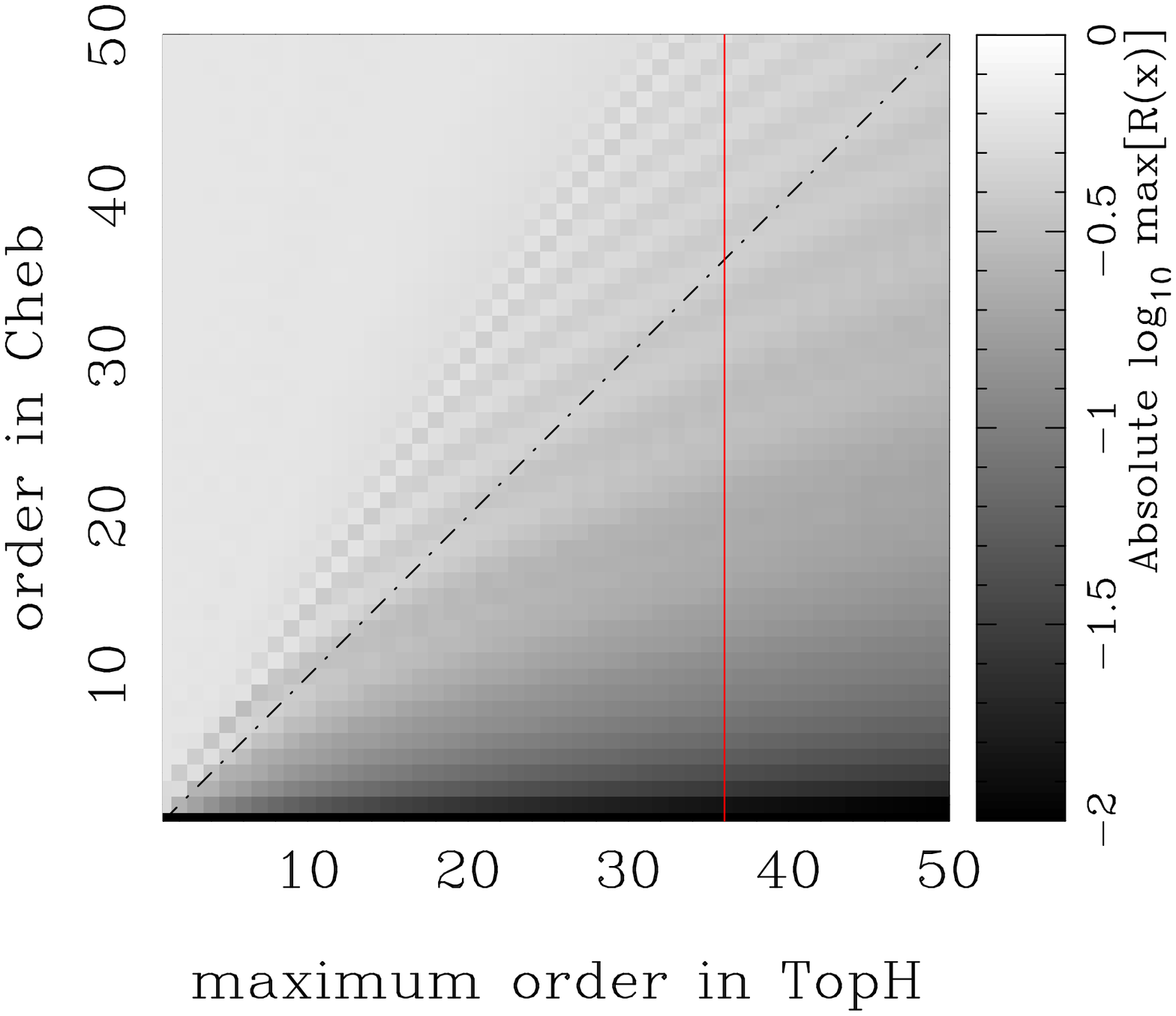, width=0.333\columnwidth, angle=0}}
\caption{An example of the residual that can occur when reconstructing a given
  basis sets functions with another basis set. On the y axes are the
  order of the basis function to be reconstructed, on the x axes are
  the maximum order of the series that is attempting to reconstruct
  the functions. The gray scale shows the maximum residual (black low
  residual, whit large residual), each function is investigated over
  the range $x \in [-1,1]$. 
  The left plot shows the order of Chebyshev
  polynomials needed to reconstruct Legendre polynomials. The middle
  plot shows the order of Legendre polynomials needed to reconstruct
  Chebyshev polynomials (the inverse operation to the left plot). The
  right hand plot shows the order of the tophat basis set needed to
  reconstruct Chebyshev polynomials. The red (gray vertical) lines are
  at $N=36$ (Albrect et al., 2009) to guide the reader.}
\label{residuals}
\end{center}
\end{figure*}

Note that interpolations between
delta function values at discrete redshifts, linearly or with some
spline function, are \emph{different} basis sets to binning. We
clarify three broad meanings of binning that occur in the literature 
\begin{itemize}
\item 
Tophat basis set : functions are not continuous, all derivatives 
are ill-defined.
\item 
Linear interpolation : functions are continuous, $1^{{\rm st}}$ derivatives are
discontinuous, $2^{{\rm nd}}$ and higher order derivatives are ill-defined.
\item 
Spline (e.g. quadratic spline) : functions are continuous, 
$1^{{\rm st}}$ derivatives are 
continuous, $2^{{\rm nd}}$ derivatives are 
discontinuous, $3^{{\rm rd}}$ derivatives and higher are ill-defined.
\end{itemize}
Of these three basis functions, only tophat form an orthogonal
basis set; though all form a complete basis set. We include
interpolation in Section \ref{Application} for completeness, and
because it is commonly used in the literature.

The most general condition that we could apply to the expected
$w(z)$ is that it is continuous (features could be very sharp in redshift 
but not 
physically discontinuous -- the sound speed of dark energy
is generally expected to be $\leq c$)\footnote{There are some tachyonic 
dark energy models (e.g. Bagla et al., 2003) but choosing a basis set
based on these specific models would be optimistic.}. 
We note that binning is not discrete differentiable
since at the bin boundary the gradient is infinite. 

We will now investigate how the Fisher matrix should be treated when
such an expanded basis set is used. 

\subsection{Basis set decomposition}
\label{Basis Set Decomposition}

The general framework with which we are now presented consists of a
Fisher matrix 
\be
\label{wfishf}
F=
\left( \begin{array}{cc}
 F^{\btheta\btheta} & F^{\btheta\bw(\phi)}  \\
F^{\bw(\phi)\btheta}   & F^{\bw(\phi)\bw(\phi)} \\
  \end{array}\right),
\ee
where $\btheta$ are the cosmological parameters not associated with
$w(z)$. $\bw(\phi)$ are the set of parameters that describe $w(z)$ where
we have expanded $w(z)$ using a general complete orthogonal basis set
with basis functions $\phi$. The sub-Fisher matrix for $w(z)$ is
calculated by 
\be 
F^{\bw(\phi)\bw(\phi)}=\sum_x \frac{1}{\sigma(x)^2}
    \frac{\partial S}{\partial a_i}\frac{\partial S}{\partial a_j},
\ee
where $\sigma(x)$ is the expected marginal error on the signal $S$ and the
errors on the coefficients of the expansion
(equation \ref{expand})
$\sqrt{(F^{\bw(\phi)\bw(\phi)})^{-1}_{ii}}=\Delta a_i$ (not
marginalising over non-$w(z)$ parameters $\btheta$).

A basis set is orthogonal if it satisfies the orthogonality
relationship  
\be
\label{ortho}
\int_R \phi_m(z)\phi_n(z)M(z){\rm d}z=c_m \delta^K_{mn}, 	
\ee
over a range $R$, where $M(z)$ is a weighting function, $c_m$ are
constants and $\delta^K_{mn}$ is the Kronecker delta. The
coefficients needed to construct an arbitrary function $f(z)$ using
the basis set $\phi$ are given by 
\be
a_n=\frac{1}{c_n}\int_R f(z)\phi_n(z)M(z){\rm d}z. 	
\ee 

In general we can construct the Fisher matrix (equation \ref{wfishf}) 
using an orthogonal
basis set for $w(z)$ (orthogonal with respect to a weight function
$M(z)$) but the sub-matrix
$F^{\bw(\phi)\bw(\phi)}$ will not be diagonal since $w(z)$ can, at a
minimum, only be observed through integral relations\footnote{Most
  articles in the literature (e.g. Albrecht et al., 2009; Huterer \&
  Starkman, 2003; Crittenden \& Pogosian, 2005) neglect 
  the weight function; we note that the tophat basis set is
  peculiar in that it is orthogonal with respect to any weight function.}.

A common approach is to look for the eigenfunctions of $w(z)$. 
The motivation for this is that the eigenfunctions are
thought to form an orthogonal basis 
set. These functions are found by rotating the Fisher matrix such that
it is diagonalised using the transformation 
\ba
\Lambda^{\phi}&=&Q^T F Q\nn
F&=&Q\Lambda^{\phi}Q^T, 
\ea
where Q is an orthogonal matrix, $Q^TQ=I$. 

By performing this transformation the non-$w(z)$ errors can also
affected. There a three possibilities: either $F$ is the full $w$ and
non-$w$ Fisher matrix in which case the non-$w$ parameters are clearly
affected, or $F$ is the $w$-only Fisher matrix, or $F$ can be the
marginalised $w$-only Fisher matrix. We discuss these later two options
in Section \ref{Marginalisation}.

The new Fisher matrix $\Lambda^{\phi}$ is a diagonal matrix. The $Q$
matrices are called the eigenmatrices (or vectors) which are a
special kind of Jacobian matrix in which the original matrix is
rotated such that is diagonalised (see Appendix A). In performing this operation
we have constructed a new set of basis functions, $e$, that are linear
combinations of the original basis set, $\phi$. $w(z)$ can now be
reconstructed using two equivalent forms 
\be 
1+w(z)=\sum_{n=1}^N a_n\phi_n(z) = \sum_{n=1}^N b_n e_n(z)
\ee
where $N$ is the order of the sub-Fisher matrix
$F^{\bw(\phi)\bw(\phi)}$ -- we have introduced this maximum order since
Fisher matrices have, by 
definition, a finite order. 
The new functions, $e$, can be constructed using the eigenmatrix 
\be 
\label{eee}
e_i(z)=\sum_{j=1}^N Q_{ij}\phi_j(z)
\ee
where (Appendix A)
\be 
Q_{ij}=\frac{\partial b_j}{\partial a_i}.
\ee
The new Fisher matrix has diagonal elements that are related to the
errors on the new functions' coefficients
\be 
\Lambda^{\phi}_{ij}=\sum_x \frac{1}{\sigma(x)^2}
    \frac{\partial S}{\partial b_i}\frac{\partial S}{\partial
      b_j}=\delta^K_{ij}\frac{1}{(\Delta b_i)^2}.
\ee

In performing such a rotation we have effectively created a new basis
set that is orthogonal with respect to a new weight function (equation
\ref{ortho}) that takes into account the covariance in the Fisher
matrix. 

We note that an operation of the form $F=Q\Lambda Q^T$ (or the inverse
$Q^TFQ=\Lambda$) leads to a 
unique $Q$ if $\Lambda$ and $F$ are fixed. However in general a
further operator $P$ can be applied to the new diagonal matrix 
$P\Lambda=P(Q^TFQ)$ to map a
matrix $F$ to any diagonal of the same dimension, where $P$ is also a 
diagonal. $P$ acts like a `stretching' (or compression), not a
rotating, operator. $P$ does not correspond to a change in parameters 
but a change in the errors on an eigenbasis parameter set. 
We do not consider this in the remainder of the article.

We also note that in general there exists operators that can rotate
from a large matrix to a smaller one. For example an $n\times n$
diagonal matrix $\Lambda^{\rm large}$ can be mapped to a smaller $m\times m$
diagonal matrix $\Lambda^{\rm small}$ where $m<n$ via the operator 
$\Lambda^{\rm small}_{kk}=A_{ki}\Lambda^{\rm large}_{ii}A^T_{ik}$
where $A_{ki}$ is an $n\times m$ matrix. So if the size of the
matrices are not specified then the operation $F=A(Q\Lambda
Q^T)A^T=R\Lambda R^T$ (where $R=AQ$) can map a matrix to a diagonal of
smaller dimension. This is equivalent to mapping the basis set
expansion from a particular order to one of a smaller maximum
order. Note that this is in general a one-way operation since
information is lost in the minimisation. 
In the remainder of this article we will only
consider mapping between matrices of the same dimension.

\subsection{The non-uniqueness of the eigenfunctions}
\label{The non-uniqueness of the eigenfunctions}

It is often assumed that an eigenfunction decomposition yields a set
of functions that are \emph{unique} (if they are not degenerate;
Albrecht et al., 2009;  
Crittenden \& Pogosian, 2005; Huterer \& Starkman, 2003). This statement comes
from matrix theory; the diagonalisation of a particular matrix can
only be done one way. However this does not mean that the
eigenfunctions of $w(z)$ from a finite Fisher matrix are unique, since
we can 
start with different basis sets and this choice is arbitrary. 
This break-down is due to the finite number of coefficients - so is
really a convergence statement. 

The general minimum assumption case we are dealing with is one in
which we consider two Fisher matrices calculated
by expanding $w(z)$ in two different basis sets $\phi$ and $\psi$. 

We can only transform one Fisher matrix $F^{\phi}\rightarrow F^{\psi}$
to another by changing the basis sets if   
\emph{all} the basis set $\phi$ 
functions $1\rightarrow N$ can be described using \emph{only} the
functions $1\rightarrow N$ from the set $\psi$, and vice versa. For example to 
rotate from basis $\phi$ to $\psi$ then 
\be
\label{resi1}
\psi_i(x)\equiv \sum_{j=1}^N J_{ij} \phi_j (x) \ \ \ \ \ \forall \ \ 1\leq i\leq N
\ee
needs to be true. 
However for any conceivable complete basis set for finite $N$ 
this sum will always have some residual. This residual $R_i(x)$ can be
quantified by  
\ba 
\label{resi}
\psi_i(x)=\sum_{j=1}^N J_{ij} \phi_j (x) + R_i(x)\nn
R_i(x)=\sum_{n=N+1}^{\infty} \left(\frac{1}{c^{\phi}}\int_R {\rm d}x'\psi_i(x')\phi_n(x')M^{\phi}(x')\right)\phi_n (x).
\ea

The notation we use throughout is that $Q$ are explicitly
eigenmatrices, $J$ are explicitly general 
Jacobian transformations, $R$ are residuals and $F$ are Fisher
matrices. Note however that an eigenmatrix $Q$ is a special kind of
Jacobian $J$ where the result of the transformation is a diagonal.
  
In Appendix B we show that if the residual is non-zero then the
eigenvalues are not equal. 

If the residual between any two basis sets is small, in the limit of
large $N$, then the eigenfunctions will agree to some
accuracy, and one would expect the best constrained eigenfunctions
from any basis set to converge. However for a finite $N$ the
eigenfunctions will only 
\emph{all} agree if the $N^{\rm th}$ basis function in one set can be
reproduced using only the functions $1\rightarrow N$ in other basis
set. 

Figure \ref{residuals} shows the residuals (equation \ref{resi})
between a variety of basis 
sets as a function of order (in equation \ref{resi} we replaced
$\infty$ with $10^7$ for which we find numerical convergence of the
results in this Figure since $50\ll 10^7$). 
This plot shows that in order to have a
residual of $<0.01$ when reconstructing Legendre polynomials with
Chebyshev the order of the Chebyshev needs to be at least as high as
the order of the basis function that is being reconstructed. However 
in the opposite case the low
order Chebyshev polynomials are difficult to reproduce with Legendre,
and the Chebyshev polynomials are difficult to reproduce with the
tophat basis set (binning). In fact Cheyshev functions are more
difficult to reproduce (especially at low order) because they are
bounded in the $y$ direction as well as the $x$ direction. 

If one chooses a maximum order of $N=36$ (Albrecht et al., 2009) for
example then there would 
be $>0.01$ residual in reconstructing Chebyshev, Legendre and tophat
basis sets from each other. As a result the reconstructed eigen
functions would not agree. We will investigate this further in Section
\ref{Application}.

To summarise this section 
\begin{itemize}
\item 
The eigenfunctions created using two different basis sets and a
finite order will only agree if the residual between the
reconstruction of the basis functions is zero. 
\item 
For a finite order there will always exist some eigenfunctions from a
given basis set that cannot be reproduced using another. So for finite
order the set of eigenfunctions is non-unique. 
\item 
For a large finite order we expect the best constrained eigen
functions to tend to agreement, and for this agreement to improve as
the order is increased\footnote{The full proof of this involves Weyl's
  matrix inequality and is beyond the scope of this article, we
  demonstrate convergence in Section \ref{Application}.}.
\end{itemize}

\subsection{Marginalisation}
\label{Marginalisation}

Another difficulty in constructing robust eigenfunctions is deciding
what to do with the non-$w(z)$ parameters. A common
approach is to marginalise over these other parameters and then
perform the eigenfunction decomposition. This in effect mixes all of
these parameters with the equation of state expansion.

Referring to equation (\ref{wfishf})
if we want to know the errors on expanded $w(z)$ parameters then we
can marginalise over the other parameters $\btheta$ by constructing
the Schur complement (Zhang, 2005) of the $F^{\bw(\phi)\bw(\phi)}$ sub-matrix 
\be 
\label{schr}
F^{\bw(\phi)\bw(\phi)}_{1S}=F^{\bw(\phi)\bw(\phi)}-F^{\bw(\phi)\btheta}
(F^{\btheta\btheta})^{-1}F^{\btheta\bw(\phi)}.
\ee
The new Fisher matrix $F^{\bw(\phi)\bw(\phi)}_{1S}$ will then give errors
on the $w(z)$ parameters already marginalised over the extra
parameters. 

We can now diagonalise $F^{\bw(\phi)\bw(\phi)}_{1S}$ like  
\ba
\label{d1}
{\rm diag}(F_{1S}^{\bw(\phi)\bw(\phi)})&=&(Q^{\rm
  mixed})^TF^{\bw(\phi)\bw(\phi)}Q^{\rm mixed}\nn
&-&(Q^{\rm
  mixed})^TF^{\btheta\bw(\phi)}(F^{\btheta\btheta})^{-1}F^{\bw(\phi)\btheta}Q^{\rm mixed}\nn 
&=& \Lambda^{\rm mixed}
\ea
where $\Lambda^{\rm mixed}$ is a diagonal matrix and $Q^{\rm mixed}$ is the
eigenmatrix of the Schur complement. Note that it is the sum of the
terms on the right hand side that form a diagonal not necessarily the
individual terms themselves. 

Alternatively we can diagonalise $F^{\bw(\phi)\bw(\phi)}$ without
marginalising so that $(Q^w)^TF^{\bw(\phi)\bw(\phi)}Q^w=\Lambda^w$
where $\Lambda^w$ is diagonal. In this case we can construct the new
full Fisher matrix in  
equation (\ref{fishf}) for which the $F^{\bw(\phi)\bw(\phi)}$
component is now diagonal 
\be
\label{fishf2}
F =
\left( \begin{array}{cc}
 F^{\btheta\btheta} & (Q^w)^T F^{\btheta\bw(\phi)}  \\
F^{\bw(\phi)\btheta} Q^w   & (Q^w)^T F^{\bw(\phi)\bw(\phi)} Q^w \\
  \end{array}\right).
\ee
the cross terms need to be modified so that the correlations between
$\btheta$ and $\bw(\phi)$ are correct. 
We can now write the new marginalised errors on the new diagonalised
$\bw(\phi)$ vectors like equation (\ref{schr}) so that 
\ba
\label{fdd}
F_{2S}^{\bw(\phi)\bw(\phi)}=&&(Q^w)^T F^{\bw(\phi)\bw(\phi)}Q^w\nn
 -&&(Q^w)^T F^{\btheta\bw(\phi)}(F^{\btheta\btheta})^{-1}F^{\bw(\phi)\btheta}Q^w.
\ea
Note that the $(Q^w)^TF^{\bw(\phi)\btheta}$ effectively acts like rotation on
$(F^{\btheta\btheta})^{-1}$ and that since  
the $\bw(\phi)$ are independent parameters to the $\btheta$ there is no reason to
generally expect that the  
second term is diagonal -- why should $(Q^w)^TF^{\bw(\phi)\btheta}$ be the
eigenmatrix of $(F^{\btheta\btheta})^{-1}$  
when the $\btheta$ 
are independent of the $\bw(\phi)$? Hence the 
$F_{2S}^{\bw(\phi)\bw(\phi)} \not= {\rm diag}(F_{1S}^{\bw(\phi)\bw(\phi)})$ 
 in general and the full eigenfunctions are  
not equal to the $\bw(\phi)$ eigenfunctions $Q^{\rm mixed}\not=Q^w$,
comparing equations (\ref{d1}) and (\ref{fdd}). 

If equation (\ref{d1}) is used to create eigenfunctions then the 
reconstructed functions will contain the eigenfunctions of the 
$w(z)$ parameters mixed with the eigenfunctions from the 
non-$w(z)$ parameters.


Our concern is that if the eigenfunctions are mixed then any statements on
the redshift sensitivity to $w(z)$ are dependant on the non-$w(z)$
parameterisation and the true $w(z)$-only sensitivity is masked. 
For example if flatness was assumed, if a spectral
index is included or its running, or if massive neutrinos were included
(to name a few cases) then conclusions made using the mixed eigenfunctions
will be different. 

\begin{figure}
\begin{center}
\psfig{file=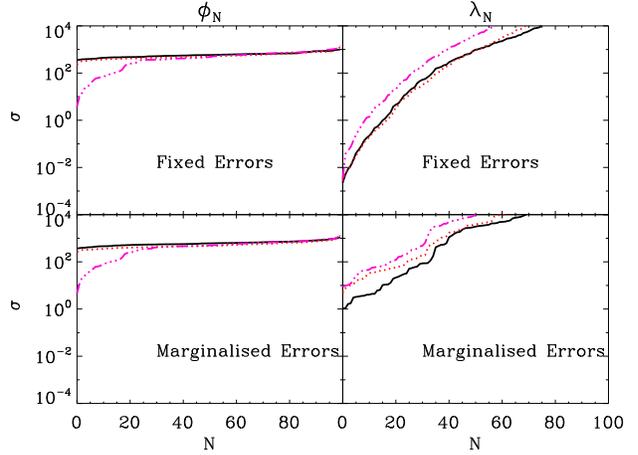, width=\columnwidth,
    angle=0}
\caption{The lines show the 
  predicted fixed errors on the basis set coefficients for the original basis
  set ${\phi_N}$ and for the rotated, eigenfunctions,
  ${\lambda_N}$ for a maximum order of $100$. We have ordered
  the errors in increasing order 
  of magnitude for clarity, so that the $N^{\rm th}$ function in the
  top panel does not corresponding to the $N^{\rm th}$ basis
  function, but is the $N^{\rm th}$ best constrained. The top two
  panels show the error for $w(z)$ not marginalised over any other
  cosmological parameters. The bottom two panels show the errors
  including marginalisation over the other non-$w(z)$ parameters using
  the method outlined in Section \ref{Marginalisation} equation (\ref{fdd}). 
  The key to the
  line styles is given in Table \ref{basissets} -- here we show the
  Fourier, Chebyshev and tophat basis sets.}
\label{errors}
\end{center}
\end{figure}

\begin{figure}
\begin{center}
\psfig{file=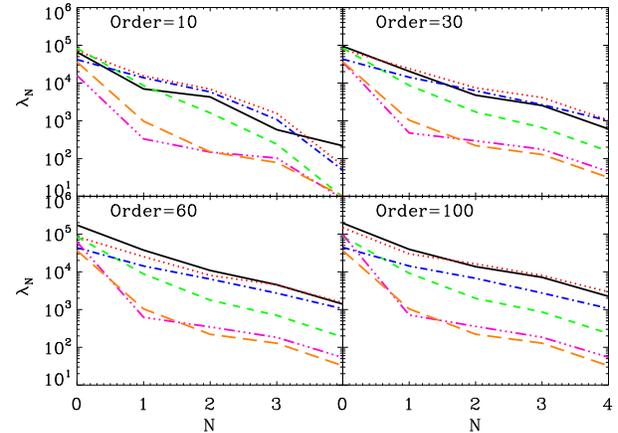, width=\columnwidth,angle=0}
\caption{The $5$ largest eigenvalues as a function of the maximum
  order of expansion. Each line represents a different basis set, 
  the key to the line styles is given in Table \ref{basissets}.}
\label{orthofig}
\end{center}
\end{figure}

\begin{figure}
\begin{center}
\psfig{file=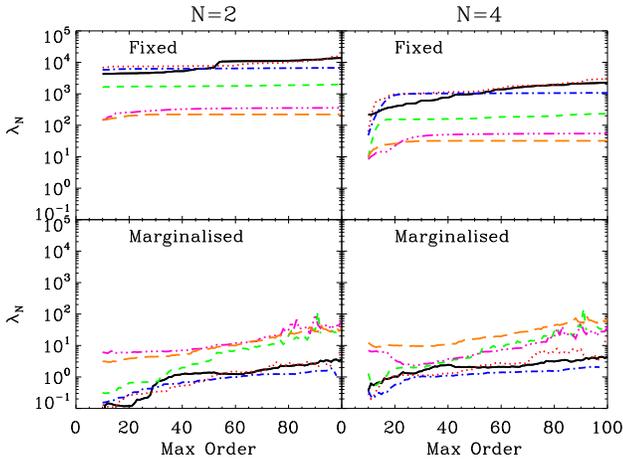, width=\columnwidth,angle=0}
\caption{The eigenvalues as a function of the maximum order in the
  basis set expansion. The key to the different basis 
  sets lines is given in Table \ref{basissets}. The upper panels show
  the eigenvalues for the second and fourth best constrained eigen
  functions and not marginalising over the non-$w(z)$ parameters. The 
  lower panels show the eigenvalues for the 
  second and fourth best constrained eigenfunction marginalising over
  the non-$w(z)$ parameters using the method outlined in Section
  \ref{Marginalisation} equation (\ref{fdd}).}  
\label{values}
\end{center}
\end{figure}

What we propose as an alternative is to find the $w(z)$-only
eigenfunctions -- which are robust to the non-$w(z)$ parameterisation
-- and to include the marginalisation in a consistent way to find the
marginalised errors on the $w(z)$-only eigenfunctions
\begin{itemize}
\item 
Find the eigenfunctions of $\bw(\phi)$ by diagonalising $F^{qq}$. 
\item 
Calculate the new full matrix like equation (\ref{fishf2}) .
\item 
Find the marginal errors on the $\bw(\phi)$ eigenfunctions by marginalising
over the other parameters like equation (\ref{fdd}).
\end{itemize}

We note that using this approach the errors on the eigenfunction  
coefficients will be correlated
after marginalising over the non-$w(z)$ parameters. But that these
marginalised eigenfunction errors will be robust to the non-$w(z)$
choices made.

\subsection{Priors}
One can add a prior before the eigendecomposition or
afterwards. Depending on the nature of the prior this can affect the
eigenfunctions themselves. Adding a general non-diagonal prior before
diagonalisation will act to modify the eigenfunctions in a similar
way as described in Section \ref{Marginalisation}. 

If we add a diagonal prior $P$ (no correlation between the prior errors)
before diagonalisation then in general this will affect the
eigenfunctions since if $QFQ^T=\Lambda_F$ then 
$QFQ^T+QPQ^T\not=$ diagonal. There is a special case of adding a unity
matrix as the prior, in this case the eigenfunctions are unaffected
since $QFQ^T+QIQ^T=QFQ^T+I=\Lambda_F+I$, and adding the prior before
or after the diagonalisation has the same effect. 

Throughout the remainder of this article we will not add any priors to
any parameters. In Albrecht et al. 2009 they suggest adding a prior of
unity to all $w(z)$ binned parameters. We note that this does not
have an effect on the eigenfunctions and in the case of survey
optimisation and comparison adds a common floor to all scenarios and
does not affect the relative merit of any survey/method.

\section{Application}
\label{Application}

We will now investigate the effect of choosing different basis sets
for the expansion of $w(z)$. We will use the example of weak lensing
tomography (described in Amara \& Refregier, 2007) where a photometric
survey is split into redshift bins and the auto and cross correlation
power spectrum of the shear fields are used to infer cosmological
parameter likelihoods. We will present predictions for a Euclid/DUNE-like
survey (Refregier, 2006) -- a $20,000$ square degree photometric
survey in $5$ bands with a number density of $35$ galaxies per square
arcminute and a median redshift of $\bar z=1$, we use $10$ tomographic
bins in the range $0 \leq z \leq 5.0$. We have extended the publicly
available \ic (Refregier et al., 2008; Kitching et al., 2009)
package to include basis set expansion of $w(z)$, these additions will
be available in version $1.2$ ({\tt http://icosmo.pbworks.com}). 

\begin{table*}
\begin{center}
\begin{tabular}{|l|c|c|c|c|}
\hline
Basis&Functions&Range $R$ &Weight $M(x)$&Plot key\\
\hline
Fourier&$\cos(nx)$ \& $\sin(nx)$&$[-\pi$,$\pi]$&$1$&$\line(1,0){10}$\\
Chebyshev&$T_n(x)=\cos(n{\rm
  a}\cos(x))$&$[-1$,$1]$&$(1-x^2)^{-\frac{1}{2}}$&
{\color{red} $\line(1,0){1}$ $\line(1,0){1}$ $\line(1,0){1}$ 
    $\line(1,0){1}$ $\line(1,0){1}$}\\
Legendre&$P_n(x)=\frac{1}{2^nn!}\frac{{\rm d}^n}{{\rm
    d}x^n}\left[(x^2-1)^n\right]$&$[-1$,$1]$&$1$&
{\color{green} $\line(1,0){4}$ $\line(1,0){4}$ $\line(1,0){4}$ 
    $\line(1,0){4}$ $\line(1,0){4}$}\\
Laguerre&$L_n(x)=\frac{{\rm e}^x}{n!}\frac{{\rm d}^n}{{\rm
    d}x^n}\left({\rm e}^{-x}x^{n}\right)$&$[0$,$\infty]$&${\rm
  e}^{-x}$&
{\color{blue} $\line(1,0){4}$ $\line(1,0){1}$ $\line(1,0){4}$ 
    $\line(1,0){1}$ $\line(1,0){4}$}\\
tophat&${\mathcal H}(x-\frac{\Delta x}{2})-{\mathcal H}(x+\frac{\Delta
  x}{2})$&$[-\infty$,$\infty]$&$1$&
{\color{pink} $\line(1,0){4}$ $\line(1,0){1}$ $\line(1,0){1}$ 
    $\line(1,0){1}$ $\line(1,0){4}$}\\
interpolation&$mx+c$&$[-\infty$,$\infty]$&$1^{*}$&
{\color{orange} $\line(1,0){8}$ $\line(1,0){8}$ $\line(1,0){8}$}\\
\hline
\end{tabular}
\caption{A summary of the basis sets investigated. The sets are
  complete over a range $R$ and orthogonal with respect to the weight
  function $M(z)$. Tophat is equivalent to binning. Interpolation means a
  linear interpolation between delta functions where
  $m=\frac{y_2-y_1}{x_2-x_1}$ and $c=y_1-mx_1$ for the nearest two
  delta functions $x_1\leq x < x_2$ to $x$. $^{*}$ interpolation is not 
  an orthogonal basis set, we include it here since this is a commonly 
  adopted expansion.}
\label{basissets}
\end{center}
\end{table*}

\begin{figure*}
\begin{center}
\resizebox{84mm}{!}{\psfig{file=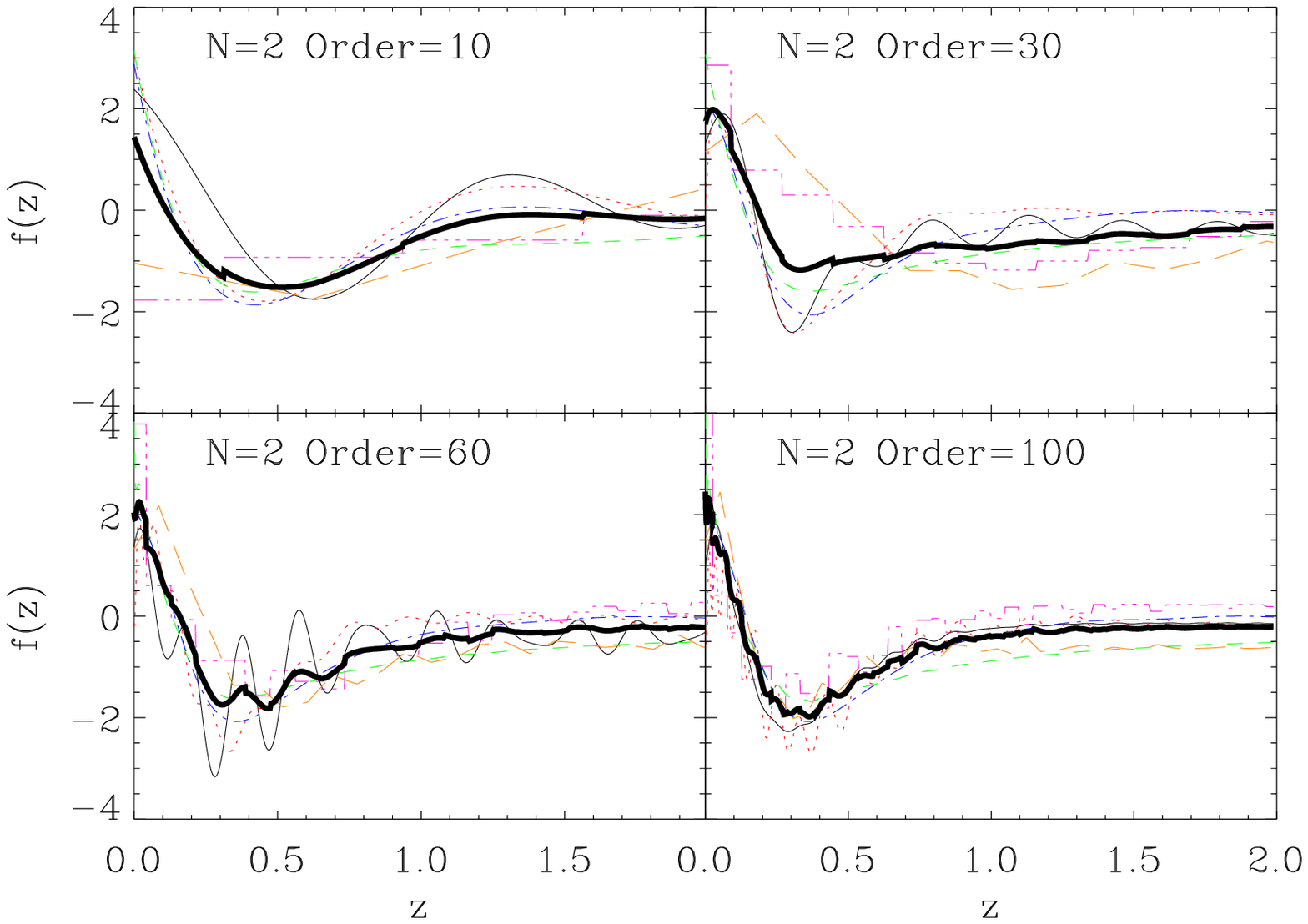, width=0.5\columnwidth,
    angle=0}}
\resizebox{84mm}{!}{\psfig{file=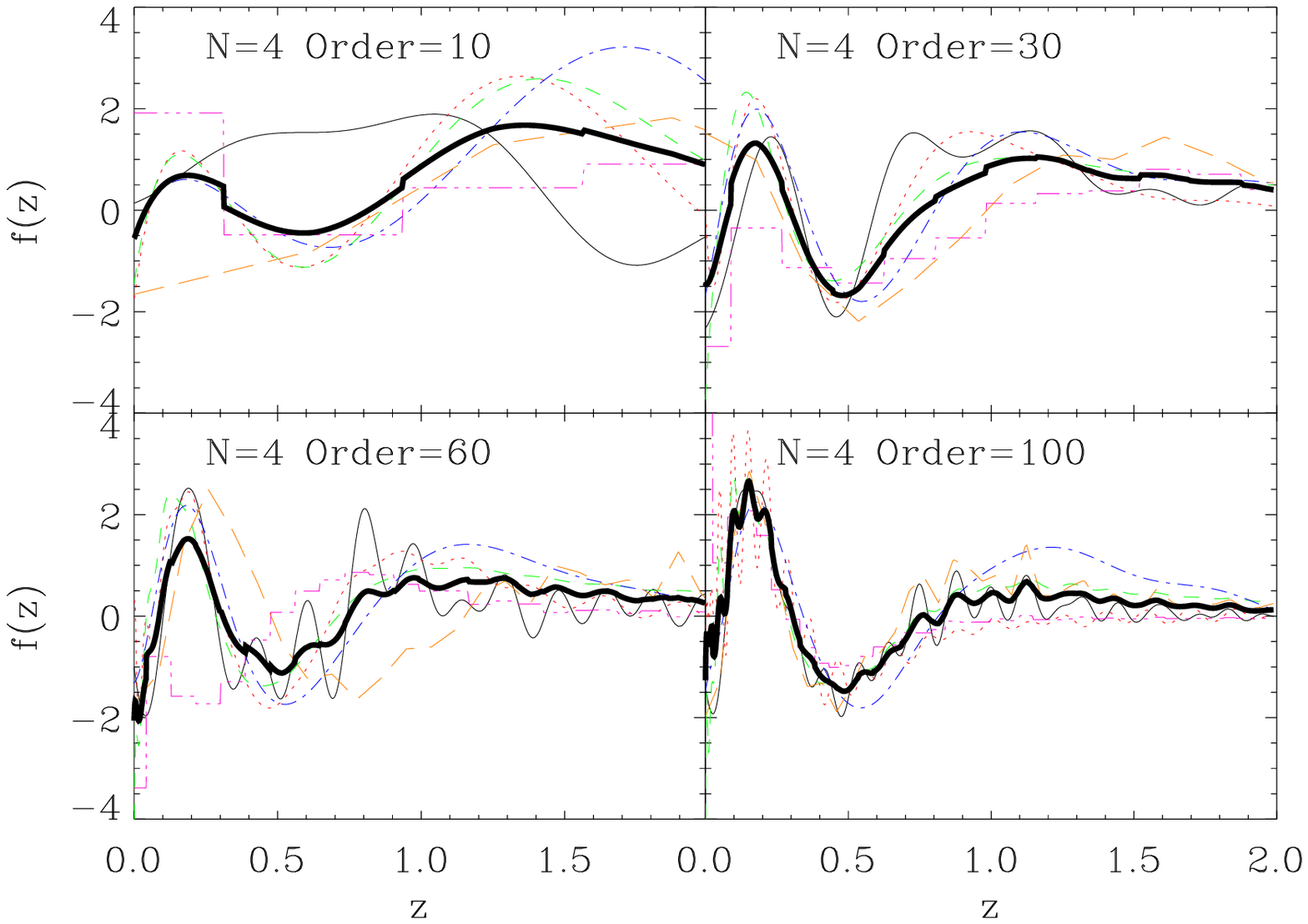, width=0.5\columnwidth,
    angle=0}} 
\caption{The second and fourth best constrained eigenfunctions for
  the weak lensing tomographic survey considered. We show how the functions
  change as the maximum order in the expansion is varied. The thin
   lines are for each individual basis set -- the key is given in
   Table \ref{basissets}. The thick black solid line shows the mean
   function over all basis sets at each order.}
\label{maxorder}
\end{center}
\end{figure*}

\begin{figure*}
\begin{center}
\resizebox{84mm}{!}{\psfig{file=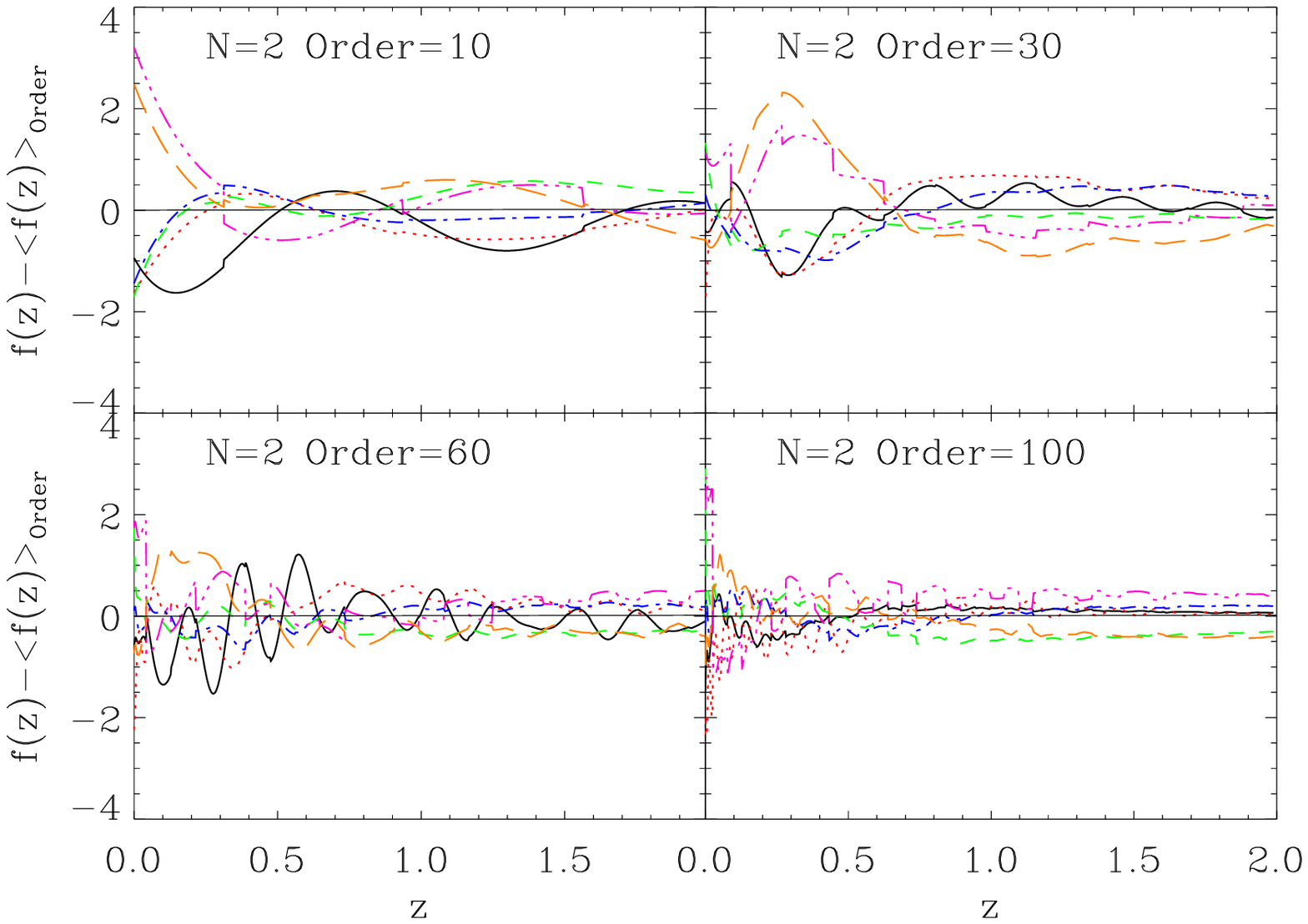, width=0.5\columnwidth,
    angle=0}}
\resizebox{84mm}{!}{\psfig{file=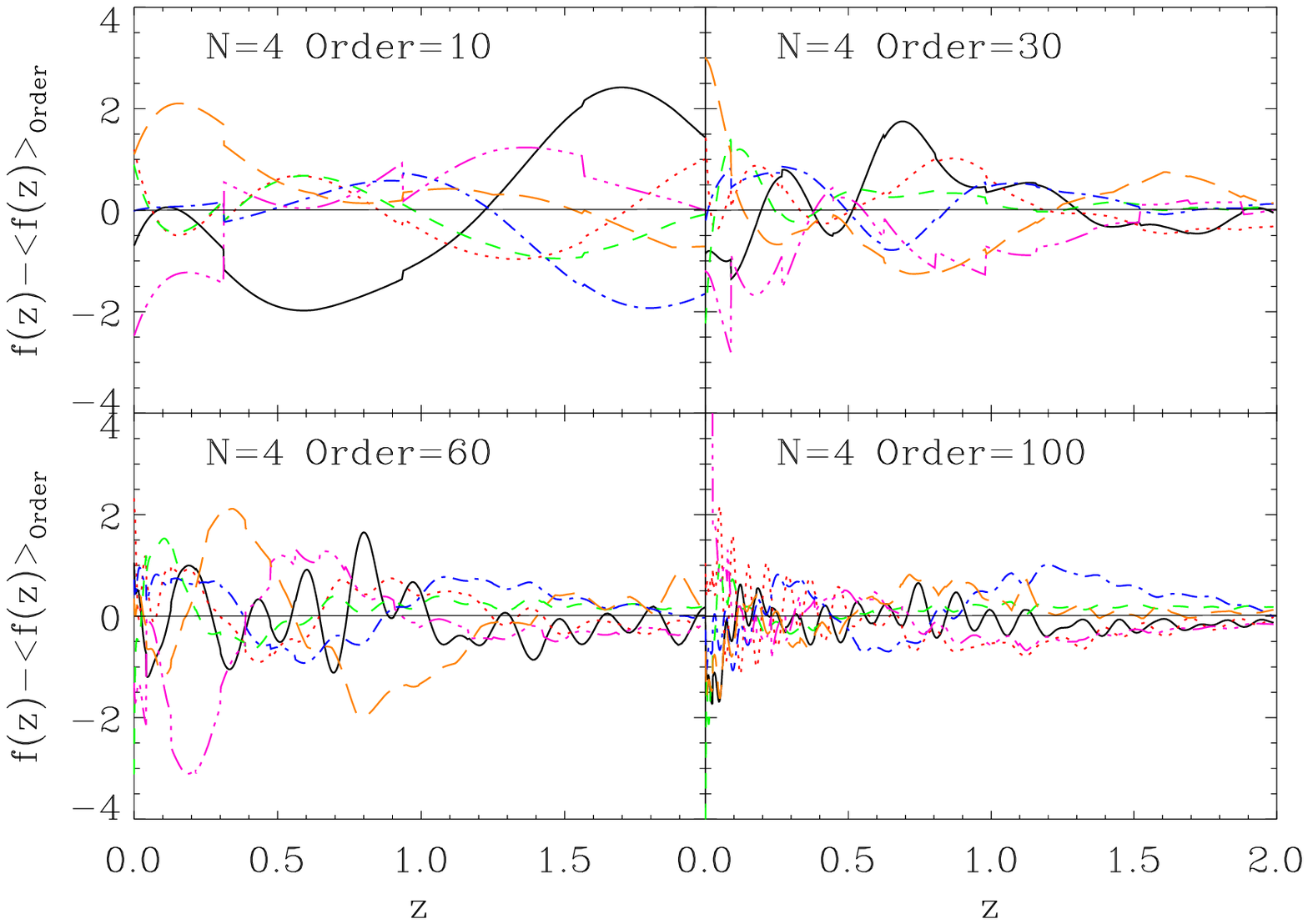, width=0.5\columnwidth,
    angle=0}} 
\caption{The difference between the eigenfunctions for each basis
set and the mean eigenfunction (over all basis sets) for each order. 
We show this
residual for the second and fourth best constrained eigenfunctions
for the weak lensing tomographic survey considered. Each panel matches
to the corresponding panel in Figure \ref{maxorder} and the key 
is given in Table \ref{basissets}.}
\label{residual}
\end{center}
\end{figure*}

We will investigate $6$ different basis sets summarised in Table
\ref{basissets}. We expand these basis sets to a maximum
order such that 
\be
1+w(z)=\sum^{\rm Order}_{N=1}a_N\phi_N(z)
\ee
where the fiducial cosmology is $\Lambda$CDM i.e. the fiducial
values are $a_N=0$ for all basis sets. In performing such a
calculation the fiducial $w(z)$ function must be one that all the
basis sets can reproduce.

\subsection{Parameter errors}

Figure \ref{errors} shows the expected $w(z)$ conditional errors on
the original basis 
function coefficients and on the eigenfunctions. It can be seen that
by rotating into an eigenbasis set the distribution of errors is
skewed to produce more well defined functions as well as more poorly
constrained functions in comparison to the non-rotated set. 

It can be seen from the left panels of Figure \ref{errors} that the
best constrained 
eigenfunctions have a similar error (the variance between the lines
is small) but for the poorest constrained functions the variance
between the basis sets is larger. This is because the most well
constrained eigenfunctions are similar for each basis set whereas the
poorest constrained have very different functional forms. As shown in
Section \ref{The non-uniqueness of the eigenfunctions} the set of
eigenfunctions is not unique. 

We also show the errors
marginalised over the non-$w(z)$ parameters. For the eigenvalue case
this is done using equation (\ref{fdd}) where the eigenfunctions are
the $w(z)$-only eigenfunctions. In all cases the best constrained
eigenfunctions are those that are also the most degenerate with the other 
cosmological parameters, and marginalising over these can increase the
error on the eigenfunctions by an order of magnitude.

Figure \ref{orthofig} shows how the largest $5$ eigenvalues change as a
function of the maximum order in the expansion. It can be seen that a
stable regime is found in which the values do not change substantially
as the order is increased. We expand this in Figure \ref{values} where
we show how the eigenvalues for the second and fourth eigenfunctions
change as a function of maximum order. The Fourier and Chebyshev
values agree at higher order, the other basis sets -- particularly
interpolations and tophat -- are very slow to converge.
Even for a maximum order of $N=100$ there
remains a large variance between the basis sets and that this variance
increases as the eigenvalues decrease. The eigenvalues are related to the 
eigenfunction errors by $\sigma=1/\sqrt{\lambda}$, we show these errors in 
Figure \ref{errors}. The best constrained $N=0$ eigenfunction has a 
similar error 
(for tophat, Chebyshev and Fourier basis sets) but by $N\geq 2$ the errors can 
vary by a factor $\gs 10$.

There is a large range of eigenvalues between the basis sets, the most
notable outliers being the tophat and interpolation 
basis sets. This is related to the fact that the residual, even at a
maximum order of $100$, between the reconstructed eigenfunctions and
the true eigenfunctions is significant. We investigate this further
in Section \ref{EigenFunctions}. When we marginalise over the
non-$w(z)$ parameters the variance between the basis sets remains but
each basis set's errors are effected in different ways since the basis
functions are degenerate with the non-$w(z)$ parameters to different
degrees. 

We have performed a numerical test that checks that the
eigendecomposition of the Fisher matrix is working correctly. First we
find 
the eigenfunctions by diagonalising the $w(z)$ sub-Fisher matrix. We
then use the eigenfunctions as a 
new basis set. We create a new Fisher matrix which takes
derivatives with respect to the coefficients of the new (eigenbasis
set). This new Fisher matrix should be diagonal and should have
diagonal elements that are equal to the original Fisher matrice's
eigenvalues. We find that the code used in this article
successfully passes this numerical check. The code used is available
as part of {\tt iCosmo} v$1.2$ and later.

\subsection{Eigenfunctions}
\label{EigenFunctions}

The eigenfunctions are reconstructed from the eigenmatrix $Q$ using
equation (\ref{eee}). Figure \ref{maxorder} shows the second and fourth
best constrained eigenfunctions for different maximum orders in the
expansion for each basis set. In the absence of any way to reconstruct
the `true' eigenmatrix we also show the mean eigenfunction
averaged over basis sets for each order -- in the limit of a high order we expect the
eigenfunctions to converge.  

It can be seen that at ${\rm Order}=40$, for example, there is still
a significant variation between basis sets. The variance between basis
sets decreases as the maximum order is increased but for the $N=4$
function the tophat basis set varies significantly from the mean. We
show the difference between each basis sets eigenfunction and the
mean in Figure \ref{residual}.

\begin{figure}
\begin{center}
\psfig{file=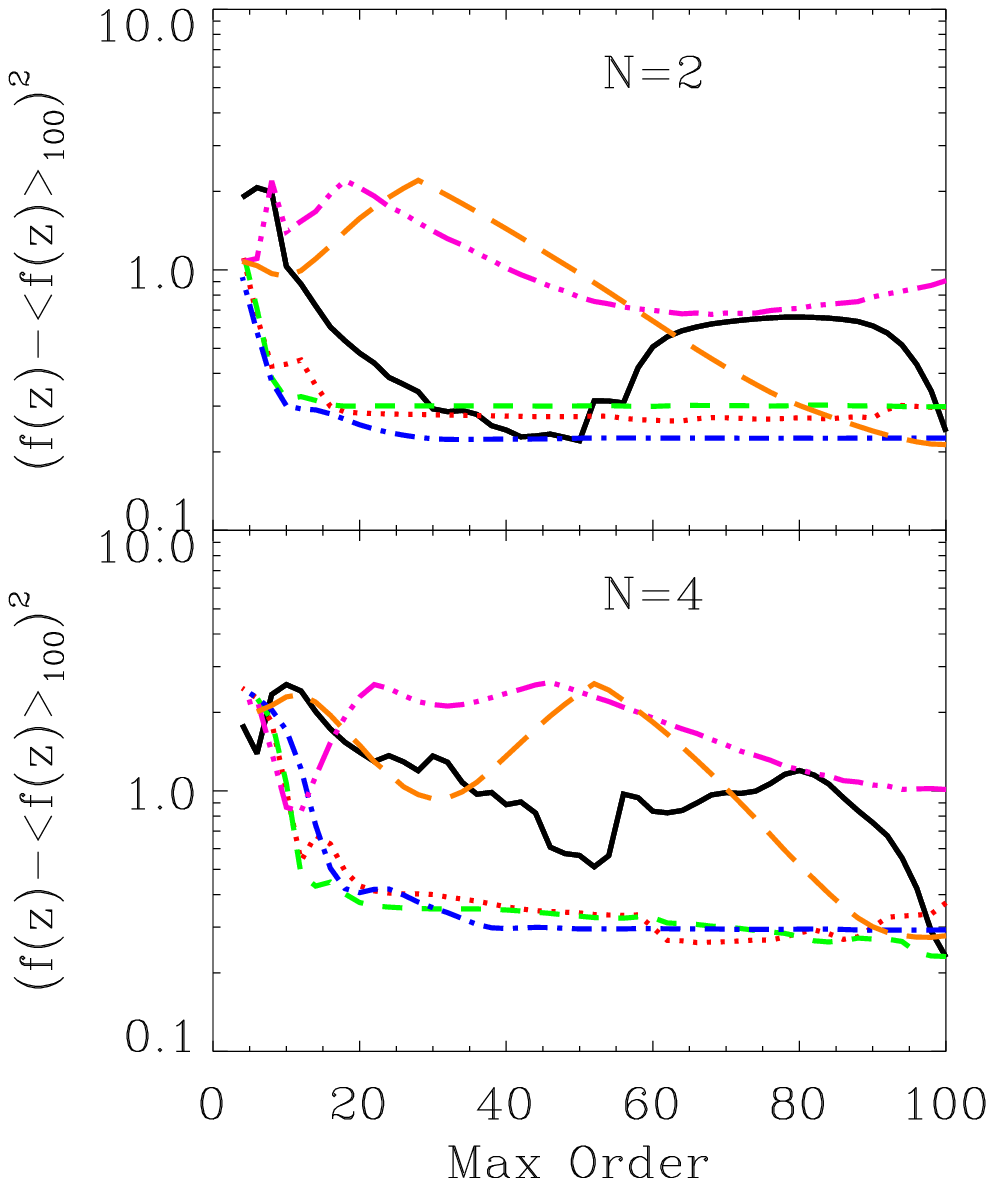, width=\columnwidth,angle=0} 
\caption{The variance between the eigenfunctions, reconstructed using
  a maximum order in the basis set expansion, and the mean eigen
  function using an order of $100$. The key to the different basis
  sets lines is given in Table \ref{basissets}. The upper panel shows
  the variance for the second best constrained eigenfunctions and the
  lower panel show the variance for the fourth best constrained eigen
  function.}
\label{variance}
\end{center}
\end{figure}

In Figure \ref{variance} we show the variance between the functions
for a given basis set and order and the mean eigenfunctions for
$N=100$. In general as the order increases the variance between the
basis sets decreases. For some basis sets such as Chebyshev, Laguerre and 
Legendre polynomials there is already
convergence even at a low order of $\sim 40$. At very low order $N\ls
10$ the eigenfunctions are 
very noisy. Other basis sets such as Fourier and interpolation reach
convergence at a high order $\sim 100$. The Fourier basis set is
noisy since high order functions can introduce highly oscillatory 
modes into the eigenfunctions, this can also be seen in the $N=60$
panel in Figure \ref{maxorder}. The tophat basis set
requires an even higher order\footnote{We have not extended this 
  calculation to even higher order since numerical effects involved in
  inverting (nearly singular) matrices of very high order start to 
  become important, we choose a safe maximum order of $100$ for this
  exercise.}.  
We note that there would be even better
agreement between the non-tophat basis sets and the mean if we excluded the
tophat basis set from this analysis.

\section{Conclusion}
\label{Conclusion}
 
We have found there is some ambiguity in the use of eigendecomposition
in the literature to date and have aimed to clarify this 
situation. By investigating an arbitrary
basis set expansion of $w(z)$ we have found that the technique of finding
the eigenfunctions of $w(z)$ from the Fisher matrix is sensitive to
the basis set choice and the order of the truncated expansion. 

We used the example of a weak lensing tomographic survey to
demonstrate how the eigenfunction decomposition is sensitive to these
choices. We find that for a Euclid/DUNE-like survey 
the best constrained eigenfunctions could be determined
to an accuracy of $\leq 1\%$ using lensing alone (no priors). 

de Putter \& Linder (2008) also
presented a critique of the eigendecomposition approach which sheds
further doubt on the technique's validity, especially at low truncated
order. They note that whilst the noise term on an eigenfunction can
be predicted it is in fact the signal-to-noise that will be important
for future surveys. We note then that a low order expansion in a
particular basis set is not robust and also has uncertain relevance in
gauging the predicted performance of a future survey.

Because of these ambiguities in the reconstruction of $w(z)$ we
recommend a move towards using physically motivated functional forms
based on models. Physical parameters can be constrained and models
compared using evidence calculations (e.g. Trotta, 2007; Heavens et
al., 2007 for a link with Fisher matrices). 

We summarise our conclusions here
\begin{itemize}
\item 
The parameterisation of $w(z)$ is an arbitrary basis set choice, and
since we do not know the nature of dark energy in general any choice is valid.
\item 
When using orthogonal basis sets it is important
to self-consistently include the weight function. 
\item 
When one refers to ``binning'' one is actually referring to a tophat
basis set. Interpolation is also valid but should not be confused with
binning. 
\item 
The set of eigenfunctions for a finite truncated basis set is not
unique. There will always exists some eigenfunctions that are not
common to all (truncated) eigenfunction sets.
\item 
When different basis sets are truncated to high order the best
constrained eigenfunctions will tend to agreement.
\item 
When marginalising over extra cosmological parameters we show that 
finding the eigenfunctions of the full Fisher matrix leads to mixed
non-$w(z)$ and $w(z)$ eigenfunctions that are dependent on the
non-$w(z)$ parameterisation. We propose an alternative in which the
marginalisation affects the $w(z)$ eigenvalues only, but not the
functions themselves, by finding the eigenfunctions of the 
$w(z)$ sub-matrix. 
\item 
To find agreement between the top $\sim 4$ eigenfunctions an order of
at least $100$ is required, for some basis set choices this is higher.  
\end{itemize}

The convergence of the eigenfunctions depends
on the original choice of basis sets -- some converge more quickly than
others. We find that Chebyshev, Legendre and Laguerre
polynomials reach convergence more quickly than interpolation, tophat
or Fourier basis sets.

Even with an order of $100$, the residuals between the reconstructed 
eigenfunctions cause the associated 
errors to vary by as much as a factor of $1$ -- $10$ 
for the best $\sim 4$ constrained functions.

We note that the dark energy Figure of Merit suggested by Albrecht et
al. (2009) uses a tophat basis set with $36$ parameters. 

We recommend that if an eigenfunction Fisher matrix analysis is
needed then a robust procedure should be followed with respect to
marginalisation such that results are not affected by non-dark energy
assumptions and priors 
and that a very high order of expansion is needed. The convergence can
be tested by investigating a variety of basis sets -- using only a
single basis set is insufficient. 

The code used in the article will form part of the \ic open
source package and will be available in version $1.2$ at {\tt
  http://www.icosmo.org}. 

\section*{Acknowledgments}
TDK is supported by STFC rolling grant number RA0888. AA is supported by
the Zwicky Fellowship at ETH Zurich. We thank Fergus Simpson, Alan
Heavens, Andy Taylor and Alexandre Refregier in particular for 
help discussions. TDK would also like to thank Sarah Bridle, 
AA would also like to thank 
Francesco Miniati and Mariano Ciccolini. We are also 
grateful to Roland de Putter for detailed comments.


\appendix

\section{Equivalence of eigen and Jacobian matrices}

In this Appendix we will show that the eigenmatrix is equal to a
Jacobian matrix. The eigenvalue decomposition of 
a matrix can be written like  
\be
\label{ev}
Q\Lambda Q^T = F.
\ee

The Jacobian transform of a
Fisher matrix maps from one parameter set $a_i$ to
another $b_i$ 
\be 
\label{jv}
J\Lambda J^T = F
\ee
where the matrix J in this case maps from the set $b$ to $a$. Where
$\Lambda$ and $F$ are defined as 
\ba
F_{ij}=\sum_x \left[\sigma^{-2}_C\frac{\partial C(\phi; x)}{\partial a_i}
\frac{\partial C(\phi; x)}{\partial a_j}\right]\nn
\Lambda_{ij}=\sum_x \left[\sigma^{-2}_C\frac{\partial C(\psi; x)}{\partial b_i}
\frac{\partial C(\psi; x)}{\partial b_j}\right]\nn
\ea
and the signal is parameterised in as 
\ba
C(\phi; x)=\sum_i a_i\phi_i(x)\nn
C(\psi; x)=\sum_i b_i\psi_i(x)
\ea
in $F$ and $\Lambda$ respectively. The elements of the Jacobian are 
\be 
J_{ij}=\frac{\partial b_j}{\partial a_i}.
\ee

When transforming from one basis set to another we can write the
following 
\ba
a_n&=&\frac{1}{c_n}\int_R f(x)\phi_n(x)M(x){\rm d}x\nn
&=&\frac{1}{c_n}\int_R \left(\sum_m b_m\psi_m(x)\right)\phi_n(x)M(x){\rm d}x\nn	
\ea 
so that the Jacobian can be written 
\be 
\frac{\partial a_n}{\partial b_m}=\frac{1}{c^{\phi}_n}\int_R
\psi_m(x)\phi_n(x)M_{\phi}(x){\rm d}x
\ee
where the weight $M$ and constant $c$ are for the $\phi$ basis set.

If we assume that $\psi$ are eigenfunctions of the $\phi$ then they can be 
reconstructed using linear combinations of the original basis
functions
\be
\psi_i(x)=\sum_j Q_{ij}\phi_j(x).
\ee

Since the decomposition of a matrix $F$ to a diagonal matrix $\Lambda$ in
equations (\ref{ev}) and equations (\ref{jv}) is unique it can easily
be seen that the eigenmatrix is in fact a Jacobian. This can be shown
in the following way. The Jacobian is written like 
\be 
\frac{\partial a_n}{\partial b_m}=\frac{1}{c^{\phi}_n}\int_R
\psi_m(x)\phi_n(x)M_{\phi}(x){\rm d}x.
\ee
We construct new basis sets as linear combinations of the old so that 
\be 
\frac{\partial a_n}{\partial b_m}=\frac{1}{c^{\phi}_n}\int_R
Q_{mk}\phi_k(x)\phi_n(x)M_{\phi}(x){\rm d}x
\ee
this leads to 
\ba
\frac{\partial a_n}{\partial b_m}&=&\frac{1}{c^{\phi}_n}Q_{mk}\int_R
\phi_k(x)\phi_n(x)M_{\phi}(x){\rm d}x\nn
&=&\frac{1}{c^{\phi}_n}Q_{mk}\delta_{kn}c^{\phi}_n\nn
J_{mn}&=&Q_{mn}.
\ea

\section{Non-uniqueness of the eigenvalues}

Here we will show that if the residual between the reconstruction of a
basis set using another is non-zero then the eigenvalues obtained from
each set via an eigenvalue decomposition of the Fisher matrices 
are not equal.

The rotation matrix from basis $\phi$ to $\psi$ can be written  
\be 
J_{ij}=\frac{1}{c^{\phi}}\int_R{\rm d}x\psi_i(x)\phi_j(x)M^{\phi}(x).
\ee 
We now assume that the new basis functions can be written as the sum
of some linear combination of the old basis functions plus some
residual (equation \ref{resi})
\be 
J_{ij}=\frac{1}{c^{\phi}}\int_R{\rm
  d}x(J^A_{ik}\phi_k(x)+R_i(x))\phi_j(x)M^{\phi}(x) 
\ee
where the rotation matrix $J^A$ rotates the old Fisher matrix to the
new Fisher matrix assuming no residual. This can now be written like  
\ba 
J_{ij}&=&J^A_{ij}+T_{ij}\nn
T_{ij}&=&\frac{1}{c^{\phi}}\int_R{\rm d}x R_i(x)\phi_j(x)M^{\phi}(x).
\ea

In the following we will use subscripts for the basis for clarity. 
We now rotate from $\phi$ to $\psi$ then diagonalise the Fisher matrix of
$\psi$ to get $\Lambda_{\psi}$ this can be written as 
\be 
\label{B4}
J^{-1}_{\psi\Lambda_{\psi}}(J^A_{\phi\psi}+T_{\phi\psi})^{-1}F_{\phi\phi}(J^A_{\phi\psi}+T_{\phi\psi})J_{\psi\Lambda_{\psi}}=\Lambda_{\psi}. 
\ee
We can also diagonalise $F_{\phi\phi}$ directly to get $\Lambda_{\phi}$
\be 
\label{B5}
J^{-1}_{\phi\Lambda_{\phi}}F_{\phi\phi}J_{\phi\Lambda_{\phi}}=\Lambda_{\phi}.
\ee

If we assume that $\Lambda_{\psi}=\Lambda_{\phi}=\Lambda$ then
comparing equations (\ref{B4}) and (\ref{B5}) 
the following must be true 
\ba
(J^A_{\phi\psi}+T_{\phi\psi})J_{\psi\Lambda}=J_{\phi\Lambda}\nn
J^A_{\phi\psi}J_{\psi\Lambda}+T_{\phi\psi}J_{\psi\Lambda}=J_{\phi\Lambda},
\ea
since we have assumed $\Lambda_{\psi}=\Lambda_{\phi}$ the first pair of 
rotation matrices commute (however in general this is not true) 
$J_{\phi\psi}J_{\psi\Lambda}=J_{\phi\Lambda}$ and 
we have 
\be
J_{\phi\Lambda}+T_{\phi\psi}J_{\psi\Lambda}=J_{\phi\Lambda}
\ee
which leads to 
\ba
T_{\phi\psi}J_{\psi\Lambda}=0\nn
T_{\phi\psi}=0.
\ea
So if  $\Lambda_{\psi}=\Lambda_{\phi}$ then $T_{\psi\phi}=0$. 
By deduction we can then assert that if 
$T_{\psi\phi}\not=0$ then $\Lambda_{\psi}\not=\Lambda_{\phi}$ and the
eigenvalues are not unique.

\end{document}